\documentclass{article}
\usepackage{latexsym}
\usepackage{amssymb}
\usepackage{amsthm}
\usepackage{amscd}
\usepackage[mathscr]{eucal}
\usepackage{amsmath}
  \usepackage{paralist}
  \usepackage{graphics} %% add this and next lines if pictures should be in esp format
  \usepackage{epsfig} %For pictures: screened artwork should be set up with an 85 or 100 line screen
 \usepackage[colorlinks=true]{hyperref}
\hypersetup{urlcolor=blue, citecolor=red}

\newtheorem{theorem}{Theorem}[section]
\newtheorem{corollary}{Corollary}

\newtheorem{lemma}[theorem]{Lemma}
\newtheorem{proposition}{Proposition}

\theoremstyle{definition}

\author{Jos\'e Antonio Alc\'antara F\'elix\\
Simone Calogero\\[0.5cm]
Department of Applied Mathematics\\
University of Granada, Spain}
\date{}
\begin{document}

\title{On a relativistic Fokker-Planck\\ equation in kinetic theory}
\maketitle

%The abstract of your paper
\begin{abstract}
A relativistic kinetic Fokker-Planck equation that has been recently proposed in the physical literature is studied. It is shown that, in contrast to other existing relativistic models, the one considered in this paper is invariant under Lorentz transformations in the absence of friction. A similar property (invariance by Galilean transformations in the absence of friction) is verified in the non-relativistic case. In the first part of the paper some fundamental mathematical properties of the relativistic Fokker-Planck equation are established. In particular, it is proved that the model is compatible with the finite propagation speed of particles in relativity. In the second part of the paper, two non-linear relativistic mean-field models are introduced. One is obtained by coupling the relativistic Fokker-Planck equation to the Maxwell equations of electrodynamics, and is therefore of interest in plasma physics. The other mean-field model couples the Fokker-Planck dynamics to a relativistic scalar theory of gravity (the Nordstr\"om theory) and is therefore of interest in gravitational physics. In both cases the existence of steady states for all possible prescribed values of the mass is established. In the gravitational case this result is better than for the corresponding non-relativistic model, the Vlasov-Poisson-Fokker-Planck system, for which existence of steady states is known only for small mass.
\end{abstract}

%The title of your section 1
\section{Introduction}

Fokker-Planck equations provide a continuous description of stochastic particles dynamics. The most basic example is Brownian's motion, the stochastic motion of a test particle immersed in a fluid in thermodynamical equilibrium. Provided the test particle is much heavier than the molecules of the fluid, it is possible to approximate the microscopic forces acting on the test particle by two driving mechanisms: diffusion and friction.
The kinetic equation that describes the evolution of the distribution function $f$ for the test particle is the linear Fokker-Planck (or Kramers) equation~\cite{ri}:
\begin{equation}\label{FPequation}
\partial_tf+p\cdot\nabla_xf=\nabla_p\cdot\left(\beta p f + \sigma\nabla_p f\right).
\end{equation}
The distribution function $f$ is a non-negative function of the variables $(t,x,p)$, where $(x,p)$ are the phase-space coordinates (position and momentum) and $t>0$ is the time variable. We assume that the mass of the test particle is one. The positive constants $\beta,\sigma$ are the friction and diffusion parameters, respectively. The stochastic differential equations for the trajectory of the test particle associated to~\eqref{FPequation} are given by the system~\eqref{SODEclass} in Section~\ref{derivation}.

Fokker-Planck equations like~\eqref{FPequation}, or variants thereof, have several applications in different fields of physics and engineering. In astrophysics, for example, they model the effect of interstellar nebulas in a galaxy~\cite{SHS} or even dark matter~\cite{chung}. In plasma physics, Fokker-Planck equations take into account the effect of grazing close encounters among the ions (the heavy particles) and the electrons.

A questionable feature of equation~\eqref{FPequation} is that the diffusion term $\sigma\Delta_pf$  in the right hand side operates with infinite velocity: if the particles are initially distributed in a compact region of space, i.e., the initial distribution $f(0,x,p)$ is compactly supported in the variable $x$, there will be instantaneously a non-zero probability (i.e., $f>0$) to find particles everywhere in space. This property is incompatible with the well-established physical law that prevents particles from moving faster than light. Recent works in the mathematical and physical literature put forward two possible ways to eliminate this undesirable feature. One consists in replacing the classical linear diffusive (Laplace) operator with a non-linear diffusion term, as in the so-called ``relativistic" heat equation, see~\cite{FCMM}.  A mathematically simpler solution is to replace~\eqref{FPequation} with a model that is still linear and, at the same time, consistent with the relativistic mechanics of particles, where the property of finite propagation speed enters in a natural fashion. The purpose of the present article is to begin the mathematical study of one such relativistic linear models.

The physical literature abounds of proposals for what should represent the correct relativistic generalization of~\eqref{FPequation}, see for instance~\cite{ck, DH} (and~\cite{DC,DH2} for an overview and an historical background to the relativistic theory of Brownian motions). Thus the first problem to face is the choice of the relativistic Fokker-Planck equation to consider. In this paper we pick the following equation:
\begin{equation}\label{relativisticFPequation}
\partial _t f+\hat{p}\cdot\nabla_xf=\nabla_{p}\cdot \left(\beta fp+\sigma D\nabla_{p}f\right),
\end{equation}
where $\hat{p}$ is the relativistic velocity,
\[
\hat{p}=\frac{p}{\sqrt{1+|p|^2}},
\]
and $D$ is the {\it relativistic diffusion matrix} given by
\[
\quad D=\frac{I+p\otimes p}{\sqrt{1+|p|^2}}.
\]
The previous model coincides with one of the equations proposed in~\cite{DH}, namely~\cite[Eq.~(47)]{DH} and it is the subject of a recent series of papers by Haba~\cite{H1,H2,H3,H4}. In these references several generalizations of~\eqref{relativisticFPequation} are introduced, including models for massless particles, for particles with spin and models with more general friction terms\footnote{We are grateful to Prof. Haba for pointing out his work to us.}.

In Section~\ref{derivation} we justify our choice for the relativistic Fokker-Planck equation~\eqref{relativisticFPequation} by showing that it maintains  certain important physical properties satisfied by the non-relativistic model~\eqref{FPequation}.  In particular we will show that for $\beta=0$, i.e., in the absence of friction, equation~\eqref{relativisticFPequation} is Lorentz invariant. Similarly, equation~\eqref{FPequation} is invariant by Galilean transformations when $\beta=0$. (Note that in both the relativistic and non-relativistic case the friction term breaks the equivalence of inertial reference systems.)  In Section~\ref{basicproperties} we prove that the solutions of~\eqref{relativisticFPequation} enjoy some other physically and mathematically desirable properties, in particular that they behave consistently with the finite propagation speed of particles.

For the applications in astrophysics (resp. plasma physics), it is necessary to add the interaction of the particles with the self-generated gravitational (resp. electric) field. In the non-relativistic case this leads to the non-linear Vlasov-Poisson-Fokker-Planck system:
\begin{subequations}\label{VPFPsystem}
\begin{align}
&\partial_tf+p\cdot\nabla_xf-\nabla_xU\cdot\nabla_pf=\nabla_p\cdot\left( p f + \nabla_p f\right),\\
&\Delta_x U=\lambda\rho,\quad\rho(t,x)=\int_{\mathbb{R}^d}\! f(t,x,p)\,\mathrm{d}p,
\end{align}
\end{subequations}
where we set all physical constant equal to one and where $\lambda=1$ in the gravitational case, while $\lambda=-1$ in the plasma physics case. In the second part of the paper (Sections~\ref{VMFPsec}-\ref{VNFPsec}) we introduce the corresponding relativistic model. In the plasma physics case we couple the relativistic Fokker-Planck equation~\eqref{relativisticFPequation} to the Maxwell equations of electrodynamics. The resulting model is the Vlasov-Maxwell-Fokker-Planck system. Note that this model is different from the one considered in~\cite{BG, Lai,YY}, which uses the non-relativistic Fokker-Planck equation~\eqref{FPequation}. In the gravitational case we couple the Fokker-Planck dynamics to a relativistic scalar theory of gravity, the Nordstr\"om theory, which has already been used as a toy model for Einstein's theory of general relativity, see~\cite{cal,CR,CSS,ST}.  Unfortunately there are fundamental difficulties, briefly recalled at the beginning of Section~\ref{VNFPsec}, in formulating  a Fokker-Planck theory in general relativity. In this paper we prefer to avoid this issue and consider instead a toy model, which we call the Vlasov-Nordstr\"om-Fokker-Planck system.

Our main result for the Vlasov-Maxwell-Fokker-Planck and Vlasov-Nordstr\"om-Fokker-Planck systems (with an external confining potential) is the existence of steady states solutions for {\it all} possible values of the mass. We do so by variational techniques inspired by~\cite{dol}. Note that in the gravitational case our result is better than for the Vlasov-Poisson-Fokker-Planck system~\eqref{VPFPsystem}$_{\lambda=1}$, for which the existence of steady states is only known for a properly small mass~\cite{bdol}. The main advantage of the relativistic model compared to the non-relativistic one is that the energy of the Vlasov-Nordstr\"om-Fokker-Planck system is positive definite.

%The title of your section 2
\section{Derivation of the relativistic Fokker-Planck model}\label{derivation}
A common way to derive Fokker-Planck type equations is to start from a system of stochastic ordinary differential equations (SODEs). The Fokker-Planck equation is the partial differential equation satisfied by the law of the stochastic process solving the SODEs.  For instance in the case of the kinetic Fokker-Planck equation~\eqref{FPequation} the relevant SODEs are given by
\begin{equation}\label{SODEclass}
\dot{x}(t)=p(t),\quad \dot{p}(t)=-\beta p(t)+\sqrt{2}\,\sigma B(t),
\end{equation}
where $B(t)$ is the standard Brownian motion in $\mathbb{R}^d$, i.e., a centered Gaussian process with covariance $\langle B(t),B(t')\rangle=\delta(t-t')$, see~\cite{Cha, ka,ri} for details.  Following this approach to derive a relativistic Fokker-Planck equation is problematic for at least two reasons. Firstly it is not so clear how to define a ``standard" relativistic Brownian motion. Secondly, there are multiple ways to derive a Fokker-Planck equation from a system of SODEs, which lead to different partial differential equations for the law of the stochastic process. For instance, equation~\eqref{FPequation} is obtained from~\eqref{SODEclass} using It\^o's calculus, whereas a different Fokker-Planck equation would be obtained by using Stratonovich's calculus\footnote{These two difficulties are in some sense equivalent, since one can modify for instance the SODEs~\eqref{SODEclass} to end up with~\eqref{FPequation} through Stratonovich's calculus.}.  As a consequence of these ``ambiguities", there exist different models in the literature which are named ``relativistic Fokker-Planck equation", see~\cite{DC,DH2} for a review.

The purpose of this section is to justify our choice for the relativistic Fokker-Planck model which will be studied in the rest of the paper. In particular we will show that it is possible to ``derive" a relativistic Fokker-Planck equation by merely demanding that certain physical properties of the non-relativistic model be maintained in the relativistic case. We shall not refer in any moment to the SODEs for the (relativistic) stochastic process, although it will be finally observed that our equation coincides with one of the models derived in~\cite{DH,DH2} by stochastic calculus methods.

We are interested in the following two important properties of the non-relativistic Fokker-Planck equation~\eqref{FPequation}:

\begin{itemize}
\item[(NR1)] In the absence of friction, i.e. when $\beta=0$, (\ref{FPequation}) is Galilean invariant\footnote{The friction term $\nabla_{p}\cdot{(\beta pf)}$ breaks the Galilean invariance of~\eqref{FPequation}, since it corresponds to the microscopic velocity-dependent force $F=-\beta p(t)$ in~\eqref{SODEclass}.}. This means that under the change of variables
\begin{align*}
\tilde{t}=t\,, \ \tilde{x}=x-ut, \ \tilde{p}=p-u,\quad & \quad \tilde{f}(\tilde{t},\tilde{x},\tilde{p}) = f(t,x,p),
\end{align*}
$\tilde{f}$ is a solution of (\ref{FPequation})$_{\beta=0} $ if and only if $f$ is a solution, $\forall u\in\mathbb{R}^d$.
\item[(NR2)] The Maxwellian distribution function
\[
\mathscr{M}(p)=e^{-\beta|p|^2/2\sigma}
\]
is a static solution of (\ref{FPequation}). In fact, up to a multiplicative constant, it is the only global equilibrium of the equation.
\end{itemize}

We propose now a relativistic generalization of (\ref{FPequation}) by requiring that the relativistic analogues of the properties (NR1) and (NR2) hold for the new model. Precisely we require that the relativistic Fokker-Planck equation should satisfy:
\begin{itemize}
 \item [(R1)] Invariance under Lorentz transformations in the absence of friction, i.e., under the change of variables\footnote{We fix $c=1$, where $c$ is the speed of light.}
\begin{align*}
u_0 & =\sqrt{1+|u|^2}, \quad\tilde{t}= u_0t-u\cdot x\: ,\quad\tilde{x}= x-u t+\frac{u_0-1}{|u|^2}u(u\cdot x),\\
\tilde{p} & =  p-u\sqrt{1+|p|^2}+\frac{u_0-1}{|u|^2}u(u\cdot p), \quad \tilde{f}(\tilde{t},\tilde{x},\tilde{p})  = f(t,x,p),
\end{align*}
$\tilde{f}$ is a solution of the frictionless equation if and only if $f$ is a solution, $\forall u\in\mathbb{R}^d$.

\item[(R2)] The function $\mathscr{J}$ defined by
\[
	\mathscr{J}(p)=e^{-\gamma\sqrt{1+|p|^2}},
\]
	must be a static solution, for some constant $\gamma>0$. $\mathscr{J}$ is known as the J\"uttner distribution (or relativistic Maxwellian).
\end{itemize}
	
The simplest  and, in our opinion, most natural way to obtain (R1) is the following. Firstly we replace the transport term in the left hand side of (\ref{FPequation}) by its relativistic counterpart\footnote{We will adopt the Einstein convention for the sum over repeated indexes. Greek indexes go from $0$ to $d$ and Latin indexes from $1$ to $d$.}
\[
\sqrt{1+|p|^2}\,\partial_t +p\cdot\nabla_x =\sum_{\mu=0}^d p^\mu\partial _\mu=p^\mu\partial _\mu,
\]
with $p^0=\sqrt{1+|p|^2}$, $p=(p^1,\cdots,p^d)$, $\partial_0=\partial _t$ and $\partial_i=\partial_{x^i}$. Secondly the diffusive operator $\Delta_p=\nabla_p \cdot \nabla_p$ on the right side of (\ref{FPequation}) is replaced by the Laplace-Beltrami operator $\Delta_p^{(h)}$ over the Riemannian manifold $(\mathbb{R}^d,h)$, where $h$ is the hyperbolic metric, i.e., the Riemannian metric induced by the Minkowski metric over the hyperboloid $\mathfrak{H}=\{(p^0,p): p^0=\sqrt{1+|p|^2}\}$. The fact that the operator $\Delta_p^{(h)}$ is Lorentz invariant is clear, since the Lorentz transformation in the momentum variable corresponds to a translation over the hyperboloid $\mathfrak{H}$. The components of the metric $h$ in the base $\partial_{p^i}\otimes\partial_{p^j}$ of the linear space of second order covariant tensor fields on $\mathfrak{H}$ are given by
\begin{eqnarray*}
h_{ij}=\delta_{ij}-\hat{p}_i\hat{p}_j,
\end{eqnarray*}
where $p_k=\delta_{kl}p^l$ and $\hat{p}=p/p_0$ is the relativistic velocity. Note that the position of the indexes (above or below) is changed using the Euclidean metric. Let
$(h^{-1})^{ij}=\delta^{ij}+p^ip^j$  denote the inverse matrix of $h_{ij}$, i.e., $(h^{-1})^{ik} h_{kj}=\delta^{i}_j$, and denote $|h|=\mathrm{det}(h_{ij})=(1+|p|^2)^{-1}$. The action of the Laplace-Beltrami operator $\Delta_p^{(h)}$ on scalar functions is given by
\begin{equation}\label{LB}
\Delta_p^{(h)}f=\frac{1}{\sqrt{|h|}}\partial_{p^i}\left(\sqrt{|h|}(h^{-1})^{ij}\partial_{p^j}f\right).
\end{equation}
Therefore the frictionless relativistic Fokker-Planck equation is
\begin{align}
\partial _t f+\hat{p}\cdot\nabla_xf=\sigma\partial_{p^i}\left(\frac{\delta^{ij}+p^{i}p^{j}}{\sqrt{1+|p|^2}}\partial_{p^j}f\right),\label{rFP}
\end{align}
where $\sigma>0$ is the diffusion constant.

To achieve (R2) we add a friction term $\partial_{p^i}(q^i(p)f)$ to the right hand side of~\eqref{rFP} such that the current
$$ A^i=\sigma\frac{\delta^{ij}+p^{i}p^{j}}{{\sqrt{1+|p|^2}}}\partial_{p^{j}}{f}+q^if$$
vanishes for $f=\mathscr{J}$. It is straightforward to verify that this happens if and only if $q^i(p)=\gamma\sigma p^i$, leading to the following relativistic Fokker-Planck equation with friction:
\begin{eqnarray}\label{FPR}
\partial _t f+\hat{p}\cdot\nabla_xf=\partial_{p^i}\left(\beta fp^i+\sigma\frac{\delta^{ij}+p^ip^j}{\sqrt{1+|p|^2}}\partial_{p^j}f\right),
\end{eqnarray}
where $\beta=\gamma\sigma$ is the friction parameter.

Our purpose in the rest of the paper is to initiate the mathematical study of~\eqref{FPR}. Before proceeding, we modify~\eqref{FPR} in two standard ways. Firstly, we set all physical  constants to unity, i.e.,  $\beta =\sigma =\gamma =1$; our results are independent from the value of the physical constants. Moreover, in order to guarantee the existence of finite mass equilibria in the whole space, we assume that the system is subject to the action\footnote{The action of the external potential is equivalent to that of a spatially dependent friction term, which can be seen by writing~\eqref{FPRfin} in the form
\[
\partial_t f+\hat{p}\cdot\nabla_x f=\nabla_p\cdot(D\nabla_p f + f\,(p+\nabla_x V)).
\] }
 of an external confining potential $V=V(x)$, and write the equation under study in the following final form
\begin{equation}\label{FPRfin}
\partial_t f+\hat{p}\cdot\nabla_x f-\nabla_x V\cdot\nabla_p f=\nabla_p(D\nabla_p f + pf),\ t>0,\ p\in\mathbb{R}^d,\ x\in\mathbb{R}^d,
\end{equation}
where $D$ is the matrix $D^{ij}=(\delta^{ij}+p^ip^j)/\sqrt{1+|p|^2}$ (the diffusion matrix). Throughout the paper we assume $V\in C^1$ and
\begin{equation}\label{V}
e^{-V}\in L^1(\mathbb{R}^3).
\end{equation}

To conclude this section we remark that~\eqref{FPR} coincides with one of the equations proposed in~\cite{DH}, namely~\cite[Eq.~(47)]{DH}. In this reference the authors derive three different relativistic Fokker-Planck equations starting from a particular relativistic Langevin dynamics  and using the pre-, mid- and post-point rule of discretization for stochastic integrals, see also~\cite{DH2}. Equation~\eqref{FPR} is the only one, among the equations introduced in~\cite{DH}, that satisfies the properties (R1)-(R2) above.

\section{Basic properties of regular solutions}\label{basicproperties}
In this section we prove some fundamental properties of regular solutions of~\eqref{FPRfin}. By regular solution we mean that
\[
0\leq f\in C([0,\infty),L^1(\mathbb{R}^d\times\mathbb{R}^d)).
\]
Since the techniques we use are rather standard, some proofs will only be sketched.

\subsection{Cauchy problem}
We begin by sketching the proof of global existence and uniqueness to the initial value problem in the class of regular solutions.  Let $f_\mathrm{in}\in L^1$ denote the initial datum of $f$, i.e., $f_\mathrm{in}(x,p)=f(0,x,p)$.
\begin{theorem}
Given $0\leq f_\mathrm{in}\in L^1$, there exists a unique global regular solution.
\end{theorem}
\begin{proof}
Approximate the external potential by a smooth function and the initial datum by a sequence $f_{\mathrm{in},m}$ of smooth, non-negative functions with compact support. By the result proved in Appendix~\ref{global}, for each fixed $m\in\mathbb{N}$ there exists a unique  $f_m\in C([0,\infty),L^2(\mathrm{d}\nu))$, solution of~\eqref{FPRfin}, where $\mathrm{d}\nu$ is the measure $\mathrm{d}\nu=\exp(\sqrt{1+|p|^2}+V)\,\mathrm{d}(p,x)$. Moreover by standard methods (see~\cite{CLR,f,Va} for instance) one can prove the $L^1$-contraction property: $\|f_k-f_m\|_{L^1}\leq\|f_{\mathrm{in},k}-f_{\mathrm{in},m}\|_{L^1}$. Thus the sequence $f_m$ converges in $L^1$ to a regular solution. The uniqueness is also a consequence of the $L^1$-contraction property.   The non-negativity of regular solutions  can be proved by studying the evolution of a suitable regularization of $\mathrm{sign} (f)$ (see again~\cite{CLR,f,Va}).
\end{proof}

We remark that it is possible to prove global existence and uniqueness of solutions with lower regularity, see~\cite{villani} for the non-relativistic case.

In the proof of the next results it will be assumed that the solution is smooth and decays rapidly at infinity. The generalization to regular solutions is achieved by introducing first a suitable smooth positive approximation $f_\varepsilon$, for which the following calculations hold up to error terms that vanish in the limit toward a regular solution (i.e., $\varepsilon\to 0)$. We refer to~\cite{bdol} for the details of this procedure in the non-relativistic case.

\subsection{Finite propagation speed}
The first property that we want to emphasize is that equation~\eqref{FPRfin} is compatible with the finite propagation speed of particles in relativity.
\begin{proposition}\label{finitespeed}
Assume that $f_{\mathrm{in}}=0$ for $|x-x_0|\leq t_0$, where $(t_0,x_0)\in (0,\infty)\times\mathbb{R}^d$. Then $f=0$ for $(t,x)\in \Lambda(t_0,x_0)$, where
\[
\Lambda (t_0,x_0)=\{(t,x)\in[0,t_0]\times\mathbb{R}^d:|x-x_0|\leq t_0-t\}
\]
is the past light cone with vertex on $(t_0,x_0)$ and base on $t=0$. In particular, if $f_\mathrm{in}=0$ for $|x|>R$, for some $R>0$, then $f=0$ for $|x|>R+t$, for all $t>0$.
\end{proposition}
\begin{proof}
Introduce the density and the current density:
\[
\rho(t,x)=\int_{\mathbb{R}^d}\! f(t,x,p)\,\mathrm{d} p,\quad j(t,x) = \int_{\mathbb{R}^d}\!\hat{p}f(t,x,p) \,\mathrm{d}p.
\]
Clearly $|j|\leq\rho$ and the continuity equation holds: $\partial_t\rho+\nabla\cdot j=0$. The result then follows by Lemma~\ref{finitespeedlemma} in Appendix~\ref{finitespeedapp}.
\end{proof}
\subsection{Mass conservation and entropy identity}
Given a regular solution $f$, the mass is
\begin{eqnarray}
 M[f](t)=\int_{\mathbb{R}^{2d}}\!f(t,x,p) \,\mathrm{d}p\,\mathrm{d} x,\label{masa}
\end{eqnarray}
and the free energy, or (relative) entropy functional is
\begin{align}\label{fl}
\mathcal{Q}[f](t)& = \int_{\mathbb{R}^{2d}}\!f(t,x,p)\left( \sqrt{1+|p|^2}+V(x)+\log f(t,x,p)\right) \,\mathrm{d}p\,\mathrm{d} x.
\end{align}
The next proposition studies the evolution of the functionals $M,\mathcal{Q}$.
\begin{proposition}\label{lema1}
For a regular solution the following holds.
\begin{itemize}
\item[(i)] The mass is constant: $M[f]=M[f_\mathrm{in}]$.
\item[(ii)] If $\mathcal{Q}_+[f_\mathrm{in}] <\infty$, where
\begin{equation}\label{q+}
\mathcal{Q}_+[f]=\int_{\mathbb{R}^{2d}}\! f\left( \sqrt{1+|p|^2}+V(x)+\log^+ f\right)\,\mathrm{d}p\,\mathrm{d} x,
\end{equation}
$\log^+f=\max(0,\log f)$, then $f\log f \in C([0,\infty),L^{1}(\mathbb{R}^{2d}))$,
$$\int_0^t\int_{\mathbb{R}^{2d}}\! D^{ij}(p)\partial_{p^{i}}{\left(\sqrt{f/\mathscr{J}}\right)}\partial_{p^{j}}{\left(\sqrt{f/\mathscr{J}}\right)\mathscr{J}}\,\mathrm{d}p\,\mathrm{d} x\,\mathrm{d} s<\infty$$
and the entropy identity holds:
\begin{align}
\frac{\displaystyle d\mathcal{Q} }{\displaystyle dt}=-4\int_{\mathbb{R}^{2d}}\!D^{ij}(p)\partial_{p^{i}}{\left(\sqrt{f/\mathscr{J}}\right)}\partial_{p^{j}}{\left(\sqrt{f/\mathscr{J}}\right)\mathscr{J}} \,\mathrm{d}p\,\mathrm{d} x.\label{deriF}
\end{align}
\end{itemize}
\end{proposition}

\begin{proof}
Proving the conservation of mass is straightforward. As to the entropy identity~\eqref{deriF}, we begin by computing
\begin{align*}
\frac{\displaystyle d\mathcal{Q} }{\displaystyle dt}=\int_{\mathbb{R}^{2d}}\! \partial _t f\left( \sqrt{1+|p|^2}+V+\log f\right)\,\mathrm{d} p\,\mathrm{d} x\: .
\end{align*}
We define  $\partial _t f=FP[f]-T[f]$, $T=\hat{p}\cdot\nabla_x-\nabla_x V\cdot\nabla_p$. First we see that
\begin{align}
\int_{\mathbb{R}^{2d}}  \! T[f] \sqrt{1+|p|^2}\,\mathrm{d} p \,\mathrm{d} x&=\int_{\mathbb{R}^{2d}}\! \left(p\cdot\nabla_xf-\sqrt{1+|p|^2}\,\nabla_x V \cdot\nabla_pf\,\mathrm{d}p\right)\mathrm{d} x \nonumber\\
&=\int_{\mathbb{R}^{2d}}\! \nabla_{x}\cdot{\left(p f \right)}\,\mathrm{d}p \,\mathrm{d} x+\int_{\mathbb{R}^{2d}}\! \hat{p}\cdot \nabla_x V f\,\mathrm{d}p \,\mathrm{d} x\nonumber\\
&=\int_{\mathbb{R}^{2d}}\!\hat{p}\cdot\nabla_x V f \,\,\mathrm{d}p \,\mathrm{d} x.\label{f2}
\end{align}
For the integral of  $T[f]V$ we have
\begin{align}
\int_{\mathbb{R}^{2d}}\!T[f]V\,\mathrm{d}p \,\mathrm{d} x
&=\int_{\mathbb{R}^{2d}}\!(V\hat{p}\cdot\nabla_xf -V \nabla_x V\cdot\nabla_pf)\,\mathrm{d}p\,\mathrm{d} x\nonumber\\
&=-\int_{\mathbb{R}^{2d}}\!\nabla_x V\cdot \hat{p}f\,\mathrm{d}p\,\mathrm{d} x-\int_{\mathbb{R}^d}\!\nabla_{p}\cdot{\left(V\nabla_x Vf\right)}\,\mathrm{d}p\,\mathrm{d} x\nonumber\\
&=-\int_{\mathbb{R}^{2d}}\!\hat{p}\cdot \nabla_x V f\,\mathrm{d}p\,\mathrm{d} x.
\end{align}
For the integral of $T[f]\log f$ we use that for $z=(x,p)$ and $A$ a vector field such that $\nabla_{z}\cdot{A}=0$, there holds
$$ \nabla_{z}\cdot{ [A\left(f\log f-f\right)]} =A\log f\cdot \nabla_z f$$
and therefore, taking $A=(\hat{p},-\nabla_x V)$, we get
\begin{align}
\int_{\mathbb{R}^{2d}}\!T[f]\left( \log f \right)\,\mathrm{d}p \,\mathrm{d} x&=\int_{\mathbb{R}^{2d}}\! (\hat{p}\cdot\nabla_xf(\log f)
-\nabla_x V\cdot\nabla_pf(\log f))\,\mathrm{d}p\,\mathrm{d} x\nonumber\\
&=
\int_{\mathbb{R}^{2d}}\! A\log f\cdot\nabla_zf \,\mathrm{d}p\,\mathrm{d} x\nonumber\\
&=\int_{\mathbb{R}^{2d}}\!\nabla_{z}\cdot{[A(f\log f-f)]}\,\mathrm{d} p\,\mathrm{d} x=0.\label{f4}
\end{align}
Adding (\ref{f2})--(\ref{f4}), we see that
\begin{align}
\int_{\mathbb{R}^{2d}}\!T[f]\left( \sqrt{1+|p|^2}+V+\log f\right)\,\mathrm{d}p \,\mathrm{d} x=0.\label{f5}
\end{align}
Now, for the term $FP[\cdot]$ we integrate by parts and obtain
\begin{align*}
&\int_{\mathbb{R}^{2d}}\!FP[f]\left( \sqrt{1+|p|^2}+V+\log f\right)\,\mathrm{d}p \,\mathrm{d} x\\
&\qquad=    \int_{\mathbb{R}^{2d}}\!\left( \sqrt{1+|p|^2}+V+\log f\right)\nabla_p\!\cdot\!\left( fp+D\nabla_{p}f\right)\,\mathrm{d}p\,\mathrm{d} x\\
&\qquad=-\int_{\mathbb{R}^{2d}}\!\left( \hat{p}_i+\frac{1}{ f}\partial_{p^{i}}{f}\right)\left( fp^i+D^{ij}\partial_{p^j}f\right)\,\mathrm{d}p\,\mathrm{d} x\\
&\qquad= -\int_{\mathbb{R}^{2d}}\!\frac{1}{ f}\left(  f \hat{p}_i+\partial_{p^{i}}{f}\right)\left(  fp^i+D^{ij}\partial_{p^j}f\right)\,\mathrm{d}p\,\mathrm{d} x.
\end{align*}
Using
$$D^{ij}\hat{p}_i =\frac{\delta^{ij}p_i+p^{j}|p|^2}{1+|p|^2}=p^{j}$$
and
$$\partial_{p^{k}}{\left(\sqrt{f/\mathscr{J}}\right)}=\frac{1}{2\sqrt{f/\mathscr{J}}}( \mathscr{J}^{-1}f \hat{p}_k+\mathscr{J}^{-1}\partial_{p^{k}}{f}),$$
we obtain
\begin{align}\label{FPder}
&\int_{\mathbb{R}^{2d}}\!FP[f]\left( \sqrt{1+|p|^2}+V+\log f\right)\,\mathrm{d}p \,\mathrm{d} x\nonumber\\
&\qquad=-4\int_{\mathbb{R}^{2d}}\!D^{ij}(p)\partial_{p^{i}}{\left(\sqrt{f/\mathscr{J}}\right)}\partial_{p^{j}}{\left(\sqrt{f/\mathscr{J}}\right)}\mathscr{J}\,\mathrm{d}p\,\mathrm{d} x.
\end{align}
Adding (\ref{f5}) and~\eqref{FPder} concludes the proof.
\end{proof}

\subsection{Steady states}
Recall that $e^{-V}\in L^1$. It is clear that, for each $M>0$, there exists a unique regular\footnote{In Appendix~\ref{global} it is proved that the operator $L$, defined by writing the equation~\eqref{FPRfin} in the form $\partial_t f=Lf$, is hypoelliptic, provided $V\in C^\infty$. From this property one obtains that the equilibria of~\eqref{FPRfin}, which solve $Lf=0$, are automatically smooth.} static solution with mass $M$ of (\ref{FPRfin}), which is given by
\begin{subequations}\label{mest}
\begin{align}
f_0(x,p) &=  m_M(x,p)=\frac{M}{\Theta}\mathscr{J}_V(x,p),\\
\textrm{where } \mathscr{J}_V(x,p)&=e^{-\left(\sqrt{1+|p|^2} +V\right)}\textrm{ and } \Theta= \int_{\mathbb{R}^{2d}}\!\mathscr{J}_V(x,p)\,\mathrm{d}p \,\mathrm{d} x.
\end{align}
\end{subequations}

Moreover, as in the non-relativistic case, one can prove that the equilibrium solution is a minimizer of the entropy functional. To see this, we first recall the following general result proved in~\cite[Lemma~1.1]{dol}, which will also play a crucial role in the following sections.
\begin{lemma}[\cite{dol}]\label{dol1}Let us consider $\Omega \subset \mathbb{R}^d$ measurable and the functional
\begin{eqnarray}\label{H}
\mathcal{H}[g]=\int_{\Omega}\!g(y)\log g(y)\,\mathrm{d}y+\int_{\Omega}\!g(y)h(y)\,\mathrm{d}y,
\end{eqnarray}
with $g\in L^{1}(\Omega)$ non-negative such that $g(\log g)^+\in L^{1}(\Omega)$. If $h\in L^1(\Omega;g(y)dy)$ is such that $e^{-h}\in L^1(\Omega;dy)$, then $g\log g \in L^1(\Omega;dy)$ and
\begin{align*}
&\mathcal{H}[g]-\mathcal{H}[m_g]\geq\frac{1}{2}\int_{\Omega}\!\left(\sqrt{g(y)}-\sqrt{m_g(y)}\right)^2\,\mathrm{d}y,\\
&\textrm{where }\quad m_g(y)=\frac{\int_{\Omega}\!g\,\mathrm{d}y}{\int_{\Omega}\!e^{-h}\,\mathrm{d}y}e^{-h}.
\end{align*}
\end{lemma}

An immediate consequence of the previous lemma is a characterization of the minimum of $\mathcal{H}$:
\begin{corollary}\label{dolc1}With the same hypotheses of Lemma~\ref{dol1},
\begin{align*}
H(M)=\inf\left\{\mathcal{H}[g]\,:\,g\geq0,\,g\in L^{1}(\Omega),\, \int_{\Omega}\!g(y)\,\mathrm{d}y=M\right\}
\end{align*}
is bounded from below for any $M>0$ and
\begin{align*}
H(M)=\mathcal{H}[\bar{g}]=M\log\left(\frac{M}{\int_{\Omega}\!e^{-h}\,\mathrm{d}y}\right)\textrm{ with }\quad \bar{g}=M\frac{e^{-h}}{\int_{\Omega}\!e^{-h}\,\mathrm{d}y}.
\end{align*}
In fact, $\bar{g}$ is the only minimum of $\mathcal{H}(g)$.
\end{corollary}

If we take $N={2d}$, $y=(x,p)$, $\Omega=\mathbb{R}^{2d}$, $g=f$ and $h=\sqrt{1+|p|^2}+V$, we have $\mathcal{H}(g)=\mathcal{Q}(f)$ and $\bar{g}=m_M$. Thus we obtain

\begin{corollary}\label{libre}
Assume that $f\in L^{1}(\mathbb{R}^{2d})$, $f\geq0$ are such that
\begin{align*}
\mathcal{Q}_+[f]=&\int_{\mathbb{R}^{2d}}\!f\left( \sqrt{1+|p|^2}+V(x)+\log^+ f\right)\,\mathrm{d}p\,\mathrm{d} x<\infty
\end{align*}
and $e^{- V}\in L^{1}(\mathbb{R}^{d})$. Then,
\begin{align*}
\mathcal{Q}[f]-\mathcal{Q}[m_M]\geq\frac{1}{2}\int_{\mathbb{R}^{2d}}\!\left(\sqrt{f(x,p)}-\sqrt{m_M(x,p)}\right)^2\,\mathrm{d}p\,\mathrm{d} x
\end{align*}
and $m_M$ is the unique minimum of
\begin{align*}
Q(M)&=\inf\left\{\mathcal{Q}[f]\,:\,f\geq0,\,f\in L^{1}(\mathbb{R}^{2d}),\,\int_{\mathbb{R}^{2d}}\!f(x,p)\,\mathrm{d}p\,\mathrm{d} x=M,\, \mathcal{Q}_+[f]<\infty\right\}\\
&= M\log\left[\frac{M}{\int_{\mathbb{R}^{2d}}\!e^{-(\sqrt{1+|p|^2}+V)}\,\mathrm{d}p\,\mathrm{d} x}\right].
\end{align*}
\end{corollary}

In the next sections we shall generalize this result to the non-linear Vlasov-Maxwell-Fokker-Planck and Vlasov-Nordstr\"om-Fokker-Planck systems.

\section{The Vlasov-Maxwell-Fokker-Planck system}\label{VMFPsec}
In the present and next sections we consider two non-linear mean field models built on the relativistic Fokker-Planck equation~\eqref{FPRfin}. These models provide a relativistic generalization of the Vlasov-Poisson-Fokker-Planck system in the plasma physics case (present section) and in the gravitational case (next section). For simplicity we shall consider only the three dimensional case, i.e., $x,p\in\mathbb{R}^3$ (the field equations change with the dimension).

The relativistic model for plasmas is obtained by coupling the relativistic Fokker-Planck equation
\begin{subequations}\label{VMFPsystem}
  \begin{align}
  \partial _t f+\hat{p}\cdot\nabla_x f+F\cdot\nabla _pf   =  \partial_{p^i}\left( fp^i+D^{ij}\partial_{p^j}f\right),\label{VPF1}
  %\quad (t,x,p)\in [0,\infty)\times\mathbb{R}^3\times\mathbb{R}^3,
  \end{align}
for the Lorentz force field (with external potential)
\begin{align}
F:[0,\infty)\times\mathbb{R}^3\times\mathbb{R}^3\to \mathbb{R}^3, \quad F=E+\hat{p}\times B-\nabla_x V
\end{align}
 and the system of Maxwell equations given by\footnote{Up to a suitable normalization of the physical constants.}
  \begin{equation}\label{M}
  \begin{array}{lclc}
    \partial_tE  = \nabla_x\wedge{B}- j, &(i) &
\nabla_x\cdot{E}=\rho,&(ii)\\
\partial_tB=-\nabla_x\wedge{E}, &(iii)  &
\nabla_x\cdot{B}=0, &(iv)
\end{array}
  \end{equation}
with
\begin{equation}\label{sources}
\rho(t,x)=\int_{\mathbb{R}^3}\!f(t,x,p)\,\mathrm{d}p,\quad j = \int_{\mathbb{R}^3}\!\hat{p}f(t,x,p)\,\mathrm{d}p,
\end{equation}
\end{subequations}
where $E,B:[0,\infty)\times\mathbb{R}^3\to \mathbb{R}^3$ are functions of $(t,x)$ that represent the electric  and the magnetic field, respectively. Note that $(\rho,j)$ satisfies the local conservation of charge
\begin{equation}\label{conscharge}
\partial_t\rho+\nabla_x\cdot{j}=0,
\end{equation}
as a direct consequence of (\ref{VPF1}), which makes it consistent to couple the Maxwell equations and the Fokker-Planck equation.

The system (\ref{VMFPsystem}) will be called the (relativistic) Vlasov-Maxwell-Fokker-Planck system, or VMFP for short. It generalizes the Vlasov-Poisson-Fokker-Plank (VPFP) system in the plasma physics case. Therefore~(\ref{VMFPsystem}) takes into account relativistic effects in a plasma, such as the propagation of electromagnetic waves. We remark that there exist other models in the literature which are named ``Vlasov-Maxwell-Fokker-Planck", see~\cite{BG, Lai,YY}. These systems couple Maxwell's equations to the non-relativistic Fokker-Planck equation~\eqref{FPequation}.

This section continues by proving the mass conservation and the entropy identity of time-dependent solutions and the existence of steady states to VMFP. The analysis of time-dependent solutions is only formal, since there is no proof of the existence of solutions with enough regularity to which apply the argument below. We shall use the terminology ``regular solution" of VMFP in a loose sense, meaning that the solution is non-negative and  sufficiently regular to enable the following calculations.

\subsection{Formal properties of regular solutions}
The mass of regular solutions of~\eqref{VMFPsystem} is defined by~\eqref{masa}; the entropy functional is defined as
\begin{align}
\mathcal{K}[f,E,B]= \int_{\mathbb{R}^6}\!f\left( \sqrt{1+|p|^2}+V+\log f\right)\,\mathrm{d}p\,\mathrm{d} x+\frac{1}{2}\int_{\mathbb{R}^3}\!\Big(\left|E\right|^2+\left|B\right|^2\Big)\,\mathrm{d}x.\label{el}
\end{align}
\begin{proposition} \label{teovmfp}For regular solutions of (\ref{VMFPsystem}) we have:
\begin{itemize}
\item[(i)] The mass is preserved: $M(t)=const.$
\item[(ii)] The entropy functional satisfies
\begin{align}
\frac{\displaystyle d\mathcal{K} }{\displaystyle dt}=-4\int_{\mathbb{R}^6}\!D^{ij}(p)\partial_{p^{i}}{\left(\sqrt{f/\mathscr{J}}\right)}\partial_{p^{j}}{\left(\sqrt{f/\mathscr{J}}\right)}\mathscr{J}\,\mathrm{d}p\,\mathrm{d} x.\label{deriF1}
\end{align}
\item[(iii)] Let $e^{-V}\in L^1(\mathbb{R}^3)$. Regular static solutions of (\ref{VMFPsystem}) with mass $M$ verify
\begin{subequations}\label{staticVMFP}
\begin{eqnarray}
 (f_0(x,p),E_0(x),B_0(x))=(m_M(x,p),-\nabla  U(x),0),
 \end{eqnarray}
where
\begin{equation}
m_M(x,p) =  \frac{M}{\Theta}\mathscr{J}_V(x,p)e^{- U(x)}, \quad \Theta=  \int_{\mathbb{R}^6}\!e^{- U(x)}\mathscr{J}_V(x,p)\,\mathrm{d}p \,\mathrm{d} x,\label{mest1}
\end{equation}
and $U$ is a solution of
\begin{equation}
-\Delta U =  \rho,\quad
 \rho=\int_{\mathbb{R}^3}\!m_M(x,p)\,\mathrm{d}p.\label{elip1}
\end{equation}
\end{subequations}
\end{itemize}
\end{proposition}
\begin{proof}
Proving (i) is straightforward.
To achieve (ii) we write $\mathcal{K}=\mathcal{Q}+\mathcal{I}$, where $\mathcal{Q}$ is given by~\eqref{fl}. Thus
  \begin{align*}
  \frac{\displaystyle d\mathcal{K} }{\displaystyle dt}=\frac{\displaystyle d\mathcal{Q} }{\displaystyle dt}+\frac{\displaystyle d\mathcal{I} }{\displaystyle dt}=\int_{\mathbb{R}^6}\!\partial _t f\left( \sqrt{1+|p|^2}+V+\log f\right)\,\mathrm{d}p\,\mathrm{d} x +\frac{\displaystyle d\mathcal{I} }{\displaystyle dt}.
  \end{align*}
Let $\partial _t f=FP[f]-T[f]$, where in this case $F=E+\hat{p}\times B-\nabla_x V$ for $T[\cdot]$. Therefore we only need to calculate the derivative of $\mathcal{I}[E,B]$ and the part of $d\mathcal{Q}/dt$ containing the term $E+\hat{p}\times B$ in $T[\cdot]$, since the other terms from $T[\cdot]$ and $FP[\cdot]$ are the same as in the linear case, cf. Proposition \ref{lema1}.
Using (\ref{M}i) y (\ref{M}iii),  we have
\begin{align*}
\frac{\displaystyle d\mathcal{I} }{\displaystyle dt}&= \int_{\mathbb{R}^3}\!(E\cdot \partial_tE+B\cdot \partial_tB)\,\mathrm{d}x\\
\nonumber &= \int_{\mathbb{R}^3}\!(E\cdot \left(\nabla_x\wedge{B}- j\right) +B\cdot \left(-\nabla_x\wedge{E}\right))\,\mathrm{d}x \\
\nonumber&= \int_{\mathbb{R}^3}\! \big[\left(E\cdot \left(\nabla_x\wedge{B}\right)-B\cdot \left(\nabla_x\wedge{E}\right)\right)-E\cdot j\big]\,\mathrm{d}x\\
\nonumber&= \int_{\mathbb{R}^3}\!\nabla_x{\big(\left(B\times E\right)}-E\cdot j\big)\,\mathrm{d}x=-\int_{\mathbb{R}^3}\!E\cdot j\,\mathrm{d}x\: .
\end{align*}
Moreover
\begin{align*}
\int_{\mathbb{R}^6}\!\left(E+\hat{p}\times B\right)\cdot\nabla_pf \left( \log f+V\right)\,\mathrm{d}p \,\mathrm{d} x = &\int_{\mathbb{R}^6}\!\nabla_{p}\cdot{ \left[\left(E+\hat{p}\times B\right)(f\log f-f)\right]}\,\mathrm{d}p\,\mathrm{d} x\\
&+\int_{\mathbb{R}^3}\!\nabla_{p}\cdot{\left(V \left(E+\hat{p}\times B\right) f\right)}\,\mathrm{d}p\,\mathrm{d} x=0.
\end{align*}
We also have
\begin{align*}
\int_{\mathbb{R}^6}\!\sqrt{1+|p|^2}\left(E+\hat{p}\times B\right)\cdot\nabla_p f \,\mathrm{d}p \,\mathrm{d} x&=-\int_{\mathbb{R}^6}\!\hat{p}\cdot\left(E+\hat{p}\times B\right)f \,\mathrm{d}p \,\mathrm{d} x\\
&=-\int_{\mathbb{R}^3}\!E\cdot j\,\mathrm{d}x.
\end{align*}
The second equality is due to the orthogonality between $\hat{p}$ and $\hat{p}\times B$ and the definition of $j$. The proof of (ii) follows easily.
%The proof of~\eqref{L2f} is a straightforward calculation and the boundedness of $f$ in $L^2$ follows from~\eqref{L2f} by Gr\"onwall's Lemma.
As to (iii), we first notice that from  (\ref{M}iii) and since $\partial_tB_0\equiv 0$, we have $0=\partial_tB_0=-\nabla\wedge{E}_0\Rightarrow \exists\, U(x)$ such that $E_0=-\nabla U(x)$. Using (\ref{M}ii) we obtain $-\Delta U=\rho$. Moreover by~\eqref{deriF1} applied to static solutions we observe that $f_0(x,p)=\alpha(x)\mathscr{J}(p)$ for some non-negative function $\alpha=\alpha(x)$. In particular, $j=0$ (since it is the integral of an odd function) and the equations for the field $B_0$ are equivalent to $\nabla_{}\times{B}_0=\nabla\cdot{B_0}=0\Rightarrow B_0\equiv 0$. Now replacing $f_0=\alpha\mathscr{J}$, $E_0=-\nabla U$ and $B_0=0$ in (\ref{VPF1}) we obtain
\[
\hat{p}\cdot\nabla\alpha+\alpha\hat{p}\cdot\nabla(U+V)=0.
\]
It is clear that the only non-trivial regular solution of the previous equation is $\alpha=Ce^{-U-V}$, where $C$ is any positive constant. The value $C=M/d$ follows by the definition of $M$.
\end{proof}

\subsection{Existence of steady states}

In this section we prove the existence of (regular) static solutions for the system (\ref{VMFPsystem}).
In particular, we want to show that the free energy functional
\begin{align*}
\mathcal{K}[f,E, B] &= \mathcal{Q}[f]+\mathcal{I}[E,B]\\
&=\int_{\mathbb{R}^6}\!f\left( \sqrt{1+|p|^2}+V+\log f\right)\,\mathrm{d}p\,\mathrm{d} x+\frac{1}{2}\int_{\mathbb{R}^3}\!\Big(\left|E\right|^2+\left|B\right|^2\Big)\,\mathrm{d}x,
\end{align*}
subject to
\begin{align*}
\nabla\cdot{E}= \rho,\quad& \nabla\cdot{B}=0,\quad\int_{\mathbb{R}^6}\!f\,\mathrm{d}p\,\mathrm{d} x=M,
\end{align*}
attains its minimum exactly in the static solution of (\ref{VMFPsystem}) with mass $M$. The following proof generalizes the one given in~\cite[Prop.~2.2]{dol} for the VPFP system. Note that the variational problem for VMFP differs from that of VPFP studied in~\cite{dol} in two aspects. Firstly, the electromagnetic field appears as an independent variable in the entropy functional, while for VPFP the electric field is given by the convolution product of $\rho$ with $1/(4\pi|x|)$. Secondly, in the variational problem for VMFP there appear the local constraints $\nabla\cdot{E}= \rho$, $ \nabla\cdot{B}=0$. Nevertheless we will be able to reduce the problem at hand to the equivalent one for the VPFP system considered in~\cite{dol}. In particular we will show that the above minimization problem is equivalent to minimizing a reduced entropy functional $\mathcal{K}_\mathrm{red}$ that resembles the free energy in the non-relativistic case. To this purpose we use the following simple result.

\begin{lemma}\label{gradiente}
The solutions of the variational problem
\[
\inf_{h\in\mathfrak{D}}\mathcal{R}(h)=\inf_{h\in\mathfrak{D}} \int_{\mathbb{R} ^3}\!|h|^2\,\mathrm{d}x,
\]
where $\mathfrak{D}=\{h\in L^2(\mathbb{R}^3)\,:\, \nabla{h}=g\}$, $g\in L^{1}(\mathbb{R}^{3})$, are of the form $h=-\nabla U$, where  $-\Delta U=g$.
\end{lemma}
 \begin{proof}
 Let $\phi$ be a test function. The first variation of $\mathcal{R}$ evaluated on a critical point has to vanish, which implies
 \begin{align*}
 \frac{\displaystyle d }{\displaystyle dt}\mathcal{R}(h+t\phi)\left|_{t=0}\right.=&\int_{\mathbb{R}^3}\! \frac{\displaystyle d }{\displaystyle dt}|h+t\phi|^2\left|_{t=0}\right.\,\mathrm{d}x=\int_{\mathbb{R}^3}\!2h\cdot\phi\,\mathrm{d}x=0.
 \end{align*}
 In particular, we can consider test functions of the form $\phi=\nabla\wedge{v}$, which entails
 \begin{align*}
 0= \int_{\mathbb{R}^3}\!h\cdot\nabla\wedge{v}\,\mathrm{d}x=-\int_{\mathbb{R}^3}\!\nabla\wedge{h}\cdot v\,\mathrm{d}x,
 \end{align*}
for all $v\in C^\infty_c(\mathbb{R}^3)$. From here we infer that $\nabla\wedge{h}=0$ and as a consequence, there exists $U$ such that $h=-\nabla U$. Substituting this value in $\nabla\cdot h=g$ concludes the proof.
\end{proof}
Next we define
\begin{align*}
\mathcal{K}_{\mathrm{red}}(f)=\int_{\mathbb{R}^6}\!f\left( \sqrt{1+|p|^2}+\frac{1}{2}U+V+\log f\right)\,\mathrm{d}p\,\mathrm{d} x,
\end{align*}
with $-\Delta U=\rho$ and $\rho=\int_{\mathbb{R}^3}\!f\,\mathrm{d}p$.
\begin{proposition} Recall the definition~\eqref{q+} of $\mathcal{Q}_+[f]$. Let
	\begin{align*}
K(M)=\inf\Big\{\mathcal{K}[f,E,B]\,:&\,f\geq0,\,f\in L^{1}(\mathbb{R}^{3}),\, \left\|f\right\|_{L^{1}(\mathbb{R}^{2d})}=M,\,\mathcal{Q}_+[f]<\infty,\\ & (E,B)\in L^{2}(\mathbb{R}^{3}) \times L^{2}(\mathbb{R}^{3})\: ,\, \nabla\cdot{E}= \rho, \, \nabla\cdot{B}=0\Big\}
\end{align*}
 and assume that $e^{- V}\in L^{1}(\mathbb{R}^{1})$. Then,
\begin{itemize}
	\item[(i)] $K(M)=\inf\left\{\mathcal{K}_{\mathrm{red}}(f)\,:\,f\geq0,\,f\in L^{1}(\mathbb{R}^{6}),\, \left\|f\right\|_{L^{1}(\mathbb{R}^{2d})}=M,\,\mathcal{Q}_+[f]<\infty\right\}$;
\item[(ii)] $K(M)$ is bounded from below for any $M>0$;
\item[(iii)] The minimizer is unique and is given by~\eqref{staticVMFP}, i.e., $K(M)=\mathcal{K}_\mathrm{red}(m_M)$.
\end{itemize}
\end{proposition}

\begin{proof} To show (i), let $X$ denote the minimizing space and define
\begin{align*}
X_1&=\left\{ f\in L^{1}(\mathbb{R}^{6})\,:\, f\geq 0,\,\left\|f\right\|_{L^{1}(\mathbb{R}^{6})}=M,\,\mathcal{Q}_+[f]<\infty\right\} ,\\
X_2&= \left\{(E,B)\in L^{2}(\mathbb{R}^{3})\times L^{2}(\mathbb{R}^{3})\,:\, \nabla\cdot{E}= \rho, \, \nabla\cdot{B}=0\right\}.
\end{align*}
The minimum (if it exists) verifies:
\begin{align*}
K(M)=\inf_X\left\{\mathcal{K}(f,E,B)\right\}&=\inf_{X_1}\left\{\inf_{X_2}\left\{\mathcal{I}(E,B)\right\}+\mathcal{Q}(f)\right\}_{\!\!\rho=\int\!\!f\,\mathrm{d}p}\\
&=\inf_{X_1}\left\{\frac{1}{2}\int_{\mathbb{R}^3}\!|\nabla U|^2\,\mathrm{d}x+\mathcal{Q}(f)\right\}_{\!\!\rho=\int\!\! f\,\mathrm{d}p}=\inf_{X_1}\left\{\mathcal{K}_{\mathrm{red}}(f)\right\},
\end{align*}
since by Lemma  \ref{gradiente}, for $g_1=\rho$ and $g_2=0$, we have $E=-\nabla U$, $-\Delta U=\rho$ and $B=-\nabla \tilde{U}$, $-\Delta \tilde{U}=0$, which implies $\tilde{U}\equiv 0$. On the other hand, we see that
$$\frac{1}{2}\int_{\mathbb{R}^3}\!|\nabla U|^2\,\mathrm{d}x=\frac{1}{2}\int_{\mathbb{R}^3}\!-U\Delta U \,\mathrm{d}x=\frac{1}{2}\int_{\mathbb{R}^3}\!\rho U\,\mathrm{d}x=\int_{\mathbb{R}^6}\!\frac{1}{2} fU\,\mathrm{d}p\,\mathrm{d} x$$
and the original problem is therefore reduced to minimize the functional $\mathcal{K}_{\mathrm{red}}(f)$, which, up to substituting $\sqrt{1+|p|^2}$ with $|p|^2/2$, coincides with the free energy in the non-relativistic case. Thus the claims (ii) and (iii) can be established as in \cite[Prop.~2.2]{dol}.
\end{proof}

To conclude this section we remark that the existence of steady states to the VMFP system can be established also by studying directly the equation~\eqref{elip1}, as done in~\cite{gsz} for the non-relativistic case. The non-existence results proved there when $e^{-V}\notin L^1$ (see also~\cite{dol}) are  valid in the relativistic case as well.

\section{The Vlasov-Nordstr\"om-Fokker Planck system}\label{VNFPsec}
In this section we introduce yet another new model, which represents a relativistic generalization of the VPFP system in the gravitational case.  It would be desirable to obtain such a model in the framework of general relativity, since the latter is the physically correct relativistic theory of gravity (as far as we know...), but this would lead inevitably to face fundamental difficulties. In fact the consistent modeling of dissipative systems in general relativity  is not yet understood, not even at a formal level, the main reason being that the Einstein equations by themselves imply that the mass/energy/momentum of the system must be conserved\footnote{The situation is similar to what happens in electrodynamics, where the Maxwell equations alone imply the conservation of charge~\eqref{conscharge} and therefore the dynamics of the coupled matter model must be compatible with it (which is true for the relativistic Fokker-Planck equation considered in the previous section).}. To overcome this (still unresolved) fundamental issue, instead of general relativity we shall use an alternative relativistic  theory of gravity, the Nordstr\"om theory, which has already been used in the collisionless case as a toy model for the more complicated Einstein-Vlasov system~\cite{CR,ST}.  The resulting system---the Vlasov-Nordstr\"om-Fokker-Planck system---will be derived using an argument similar to the one applied in Section~\ref{derivation}.

\subsection{Derivation of the model}
While for the  VMFP system the background space-time is given by the manifold $(\mathbb{R}^4,\eta)$, where  $\eta$ is the  Minkowski metric, in the present case we assume that the space-time is given by the Lorentzian manifold $(\mathbb{R}^4,g)$, where
\begin{eqnarray*}
g=e^{2\phi}\eta,
\end{eqnarray*}
with $\phi:\mathbb{R}^4\to\mathbb{R}$ a scalar field, which will play the role of the gravitational field. Let $(t,x^1,x^2,x^3)$ be a system of coordinates which set the Minkoski metric in the canonical form $\eta_{\mu\nu}=\textrm{diag}(-1,1,1,1)$. Then
\begin{eqnarray}
g=-e^{2\phi}dt^2+e^{2\phi}\delta_{ij}\mathrm{d}x^i\mathrm{d}x^j.\label{ge}
\end{eqnarray}

The geodesics of the metric (\ref{ge}) are the solutions of the following system of ODEs:
\begin{align}
\frac{\displaystyle dt}{\displaystyle ds}=p^0 ,\quad
\frac{\displaystyle dx^i}{\displaystyle ds}=p^i, \quad
\frac{\displaystyle dp^\mu}{\displaystyle ds}=-\Gamma^\mu_{\nu\sigma}p^\nu p^\sigma ,\label{geo}
\end{align}
where $s$ is the geodesic parameter and $\Gamma^\mu_{\nu\sigma}$ are the Christoffel symbols of $g$:
\begin{equation}
\Gamma^\mu_{\nu\sigma}=\frac{1}{2} g^{\mu\gamma}\left(\partial_\nu g_{\sigma\gamma}+\partial_\sigma g_{\nu\gamma}-\partial_\gamma g_{\nu\sigma}\right)=\delta ^\mu _\nu \partial_\sigma\phi+\delta ^\mu _\sigma\partial_\nu\phi - e^{-2\phi}g_{\nu\sigma}\eta^{\mu\gamma}\partial_\gamma\phi\:.\label{sim}
\end{equation}

Let us consider a system of particles with unit mass that move along the geodesic curves. The geodesic motion reflects the physical property that the particles interact only through the gravitational field. If we want to interpret $p^\mu$ as the four-momentum of the particles, we need to impose that $p^\mu$ has length equal to $-1$, i.e., $g_{\mu\nu}p^\mu p^\nu=-1$.
This entails
\begin{eqnarray}
p^0=\sqrt{e^{-2\phi}+|p|^2},\quad|p|^2=\delta_{ij}p^ip^j.\label{pcero}
\end{eqnarray}

Let $f(t,x,p)$, $x=(x^1,x^2,x^3)$ and $p=(p^1,p^2,p^3)$, be the distribution function of particles in the position $x$ at time $t$ and with four-momentum $p^\mu=(p^0,p)=(\sqrt{e^{-2\phi}+|p|^2},p)$. Having assumed that the solutions of (\ref{geo}) are the particles trajectories, we obtain that $f$ satisfies the equation
\begin{eqnarray*}
p^0\partial_tf+p\cdot\nabla_xf-\Gamma^i_{\mu\nu}p^\mu p^\nu\partial_{p^i}f=0,
\end{eqnarray*}
where $p^0$ is given by (\ref{pcero}). Substituting (\ref{sim}) in the last equation we obtain
\begin{eqnarray*}
p^0\partial_tf+p\cdot\nabla_xf-\left[2\left(p_0\partial_t\phi+p\cdot\nabla_x\phi\right)p+e^{-2\phi}\nabla_x\phi\right]\cdot\nabla_{p}f=0.
\end{eqnarray*}
The previous equation is the Vlasov equation for collisionless particles. For the Fokker-Planck equation we need to add a diffusion and a friction term in the right hand side. Motivated by the discussion in Section~\ref{derivation}, for the diffusion term we pick $\Delta^{(h)}_pf$, where  $h$ is the metric induced by (\ref{ge}) over the hyperboloid $p^0=\sqrt{e^{-2\phi}+|p|^2}$. It can be verified that\footnote{Although the metric $g$ is not Euclidean, we keep using the metric  $\delta_{ij}$ for moving up and down indexes.}
\[
h_{ij}=e^{2\phi}\left(\delta_{ij}-\frac{p_ip_j}{e^{-2\phi}+|p|^2}\right),\quad p_i=\delta_{ij}p^j.
\]
 We have
\begin{align*}
|h|=\det h=\left(e^{-2\phi}+|p|^2\right)^{-1}e^{4\phi},\quad
(h^{-1})^{ij}=e^{-2\phi}\delta^{ij}+p^ip^j.
\end{align*}
Therefore,
\begin{align*}
\Delta^{(h)}_pf&=\frac{1}{\sqrt{|h|}}\partial_{p^{i}} \left(\sqrt{|h|}(h^{-1})^{ij}\partial_{p^{j}}f\right)\\
&=\sqrt{e^{-2\phi}+|p|^2}\,\partial_{p^{i}}\left(\frac{e^{-2\phi}\delta^{ij}+p^ip^j}{\sqrt{e^{-2\phi}+|p|^2}}\partial_{p^{j}}f\right).
\end{align*}
We then obtain the Fokker-Planck equation in the absence of friction in the following form:
\begin{subequations}\label{temporal}
\begin{equation}
Sf-\left[2S\phi\, p+\frac{e^{-2\phi}\nabla_x\phi}{\sqrt{e^{-2\phi}+|p|^2}}\right]\cdot\nabla_{p}f=\partial_{p^{i}}\left(\frac{e^{-2\phi}\delta^{ij}+p^ip^j}{\sqrt{e^{-2\phi}+|p|^2}}\partial_{p^{j}}f\right),
\end{equation}
where
\begin{equation}
S u= \partial_t u+\frac{p}{\sqrt{e^{-2\phi}+|p|^2}}\cdot\nabla_xu.
\end{equation}
For the scalar gravitational field $\phi$ we postulate the non-linear wave equation
\begin{eqnarray}
\Box \phi:=\partial^2_t\phi-\Delta _x\phi=-e^{6\phi}\int_{\mathbb{R}^3}\!\frac{f(t,x,p)}{\sqrt{e^{-2\phi}+|p|^2}}\,\mathrm{d}p,
\end{eqnarray}
\end{subequations}
which has been justified in~\cite{cal}. Now, doing the change of variables  $\tilde{f}(t,x,p)=f(t,x,e^{-2\phi}p)$, the system~\eqref{temporal} takes the form
\begin{subequations}\label{VNFPnofriction}
\begin{align}
&\partial_t\tilde{f}+\nabla_p\left(\sqrt{e^{2\phi}+|p|^2}\right)\cdot\nabla_x\tilde{f}-\nabla_x\left(\sqrt{e^{2\phi}+|p|^2}\right)\cdot\nabla_p\tilde{f}=\partial_{p^{i}}\left(\Lambda_\phi^{ij}(p)\partial_{p^{j}}\tilde{f}\right),\label{nvfps}\\
&\Box \phi=-e^{2\phi}\int_{\mathbb{R}^3}\!\frac{\tilde{f}(t,x,p)}{\sqrt{e^{2\phi}+|p|^2}}\,\mathrm{d}p,\label{econdas}
\end{align}
where
\begin{equation}\label{lambdadef}
\Lambda_\phi^{ij}(p)=\frac{e^{4\phi}\delta^{ij}+e^{2\phi}p^{i}p^{j}}{\sqrt{e^{2\phi}+|p|^2}}.
\newcommand{\ju}{e^{-\sqrt{1+|p|^2}}}
\end{equation}
\end{subequations}
The system~\eqref{VNFPnofriction} is the Vlasov-Nordstr\"om-Fokker-Planck system in the absence of friction. It is invariant under the Lorentz type transformations given in~\cite{CCSS}. To introduce a friction term, we first notice that for any given time independent scalar function $\phi_0=\phi_0(x)$, the left hand side of (\ref{nvfps})  vanishes for
 \begin{eqnarray}
\tilde{f}=\tilde{f}_0(x,p)=e^{-\sqrt{e^{2\phi_0}+|p|^2}}.\label{jutn}
\end{eqnarray}
This suggests to introduce a friction term of the form $\nabla_{p}\cdot({q\tilde{f}})$  on the right side of~\eqref{nvfps} such that
\[
\Lambda_\phi^{ij}(p)\partial_{p^j}\tilde{f}+q^i\tilde{f}=0,\quad\text{if }\ \tilde{f}=e^{-\sqrt{e^{2\phi}+|p|^2}}.
\]
 It can be verified that $q=e^{2\phi}p$. Adding this friction term and an external potential to (\ref{nvfps}), we get
\begin{subequations}\label{vnfp}
\begin{align}
\partial_tf+\nabla_p\left(\sqrt{e^{2\phi}+|p|^2}\right)\cdot\nabla_xf&-\nabla_x\left(\sqrt{e^{2\phi}+|p|^2}+V(x)\right)\cdot\nabla_pf\label{vfpn}\\
&=\partial_{p^{i}}\left(\Lambda_\phi^{ij}(p)\partial_{p^{j}}f+e^{2\phi}p^if\right),
\nonumber\\
\Box \phi=-e^{2\phi}\int_{\mathbb{R}^3}\!\frac{f(t,x,p)}{\sqrt{e^{2\phi}+|p|^2}}\,\mathrm{d}p,\label{nos}
\end{align}
where $\Lambda_\phi^{ij}(p)$ is given by~\eqref{lambdadef} and where we removed the tilde for notational simplicity.
\end{subequations}
The system (\ref{vnfp}) will be called the Vlasov-Nordstr\"om-Fokker-Planck (VNFP) system.

\subsection{Formal properties of regular solutions}
Besides regularity, we assume that the solutions of VNFP are such that $e^\phi$ is bounded in any finite interval of time. This is true as soon as the initial data for the field equation~\eqref{nos} are bounded. To see this, note that regular solutions of~\eqref{nos} verify $\phi=\phi_{\mathrm{hom}}+\psi$, where $\phi_\mathrm{hom}$ solves the wave equation $\Box\phi_\mathrm{hom}=0$ with the same data of $\phi$ and $\psi$ solves~\eqref{nos} with zero data. Since the right hand side of~\eqref{nos} is non-negative, then $\psi\leq 0$, and therefore $e^{\phi}=e^{\phi_\mathrm{hom}}e^\psi\leq e^{\phi_\mathrm{hom}}$ is bounded, as we claimed.

The mass of regular solutions of VNFP is defined by~\eqref{masa}. The entropy functional is
\begin{align}
\mathcal{K}[f,\phi,\partial_t\phi] =& \int_{\mathbb{R}^6}\!f\left( \sqrt{e^{2\phi}+|p|^2}+V(x)+\log f\right)\,\mathrm{d}p\,\mathrm{d} x\nonumber\\
&+\frac{1}{2}\int_{\mathbb{R}^3}\!\Big(\left|\partial_t\phi\right|^2+\left|\nabla_x\phi\right|^2\Big)\,\mathrm{d}x\nonumber\\
=&\ \mathcal{Q}[f,\phi]+\mathcal{I}[\phi,\partial_t\phi].\label{eln}
\end{align}
Note that the energy part of the entropy functional is positive definite, in contrast to the case of the gravitational VPFP system.
\begin{proposition} For regular solutions of (\ref{vnfp}), we have:
\begin{itemize}
\item[(i)] $M(t)\equiv$constant.
\item[(ii)] The entropy functional
satisfies
\begin{align}
 \frac{\displaystyle d\mathcal{K} }{\displaystyle dt}=-4\frac{}{ }\int_{\mathbb{R}^6}\!\Lambda_\phi^{ij}(p)\partial_{p^{i}}{\left(\sqrt{f/\mathscr{J}^\phi}\right)}\partial_{p^{j}}{\left(\sqrt{f/\mathscr{J}^\phi}\right)}\mathscr{J}^\phi\,\mathrm{d}p\,\mathrm{d} x,\label{deriF1n}
\end{align}
where $\mathscr{J}^\phi(x,p)=e^{-\sqrt{e^{2\phi}+|p|^2}}$.
\item[(iii)] Let $e^{-V}\in L^1$. Static solutions of VNFP with mass $M>0$ are of the form
\begin{subequations}\label{staticVNFP}
\begin{equation}
(f_0(x,p),\phi_0(x))=(m_M(x,p),\phi_0(x)),
\end{equation}
where\footnote{Note that $\mathscr{J}_V\in L^1(\mathbb{R}^6)$, because $e^{\phi_0}$ is bounded.}
\begin{equation}\label{staticVNFP2}
m_M(x,p)=\frac{M}{\Theta}\mathscr{J}_V(x,p),\quad\mathscr{J}_V= e^{-\sqrt{e^{2\phi_0}+|p|^2}-V},\quad \Theta=\int_{\mathbb{R} ^6}\!\mathscr{J}_V(x,p)\,\mathrm{d}p\,\mathrm{d} x,
\end{equation}
and  $\phi_0$ solves
\begin{equation}\label{eliN}
\Delta \phi_0 =e^{2\phi_0}\int_{\mathbb{R}^3}\!\frac{m_M(x,p)}{\sqrt{e^{2\phi_0}+|p|^2}}\,\mathrm{d}p.
\end{equation}
\end{subequations}
\end{itemize}
\end{proposition}

\begin{proof}
The proof of (i) is straightforward.
To show (ii), we first observe that
\begin{equation}\label{tempica}
\frac{d\mathcal{Q}}{dt}=\int_{\mathbb{R}^6}\!\partial_t f\left( \sqrt{e^{2\phi}+|p|^2}+V+\log f\right)\mathrm{d}p\,\mathrm{d}x+\int_{\mathbb{R}^6}\!\frac{f e^{2\phi}\partial_t\phi}{\sqrt{e^{2\phi}+|p|^2}}\,\mathrm{d}p\,\mathrm{d} x.
\end{equation}
Again we split $\partial_tf=FP[f]-T[f]$, where $T=\nabla_x(\sqrt{e^{2\phi}+|p|^2}-V)\cdot\nabla_p-\nabla_p (\sqrt{e^{2\phi}+|p|^2})\cdot\nabla_x$. For the integral containing $T$ we have
 \begin{align*}
 &\int_{\mathbb{R}^6}\!T[f]\sqrt{e^{2\phi}+|p|^2}\,\mathrm{d}p\,\mathrm{d} x=-\int_{\mathbb{R}^6}\!\frac{p\cdot\nabla_x V}{\sqrt{e^{2\phi}+|p|^2}}\,\mathrm{d}p\,\mathrm{d} x,\\
 &\int_{\mathbb{R}^6}\!T[f]\log f\,\mathrm{d}p\,\mathrm{d} x=0,\\
 &\int_{\mathbb{R}^6}\!T[f]V\,\mathrm{d}p\,\mathrm{d} x=\int_{\mathbb{R}^6}\!\frac{p\cdot\nabla_x V}{\sqrt{e^{2\phi}+|p|^2}}\,\mathrm{d}p\,\mathrm{d} x.
\end{align*}
Thus the term involving $T[f]$ gives no contribution.
Moreover
  \begin{align*}
 & \int_{\mathbb{R}^6}\!FP[f]\left( \sqrt{e^{2\phi}+|p|^2}+V+\log f\right)\,\mathrm{d}p\,\mathrm{d} x\\
   &\qquad=
\int_{\mathbb{R}^{6}}\!\left( \hat{p}_i+\frac{1}{ f}\partial_{p^{i}}{f}\right)\left( e^{2\phi}fp^i+\Lambda_\phi^{ij}(p)\partial_{p^j}f\right)\,\mathrm{d}p\,\mathrm{d} x\\
&\qquad=-4\int_{\mathbb{R}^6}\!\mathscr{J}^\phi\Lambda_\phi^{ij}(p)\partial_{p^{i}}{\left(\sqrt{f/\mathscr{J}^\phi}\right)}\partial_{p^{j}}{\left(\sqrt{f/\mathscr{J}^\phi}\right)}\,\mathrm{d}p\,\mathrm{d} x,
\end{align*}
where we used that
\begin{align*}
e^{2\phi}p^j&=\Lambda_\phi^{ij}(p) \hat{p}_i\ \text{ and }\
\partial_{p^{k}}{\left(\sqrt{f/\mathscr{J}^\phi}\right)}=\frac{(\mathscr{J}^\phi)^{-1}}{2\sqrt{f/\mathscr{J}^\phi}}\left(  f \hat{p}_k+\partial_{p^k}{f}\right).
\end{align*}
On the other hand,
\[
\frac{d\mathcal{I}}{dt}=\int_{\mathbb{R}^3}\!\big(\partial_t\phi\partial_t^2\phi+\nabla_x\phi\cdot\nabla\partial_t\phi\big)\,\mathrm{d}x=\int_{\mathbb{R}^3}\!\partial_t\phi\Box\phi\,\mathrm{d}x,
\]
which cancels the last term in~\eqref{tempica} due to the field equation~\eqref{nos}. This concludes the proof of (ii).
For the last statement, we use that, by (ii), static solutions must have the form $f_0(x,p)=\alpha(x)\mathscr{J}^{\phi_0}(x,p)$. Substituting in (\ref{vfpn}) we obtain the equation $p\cdot(\nabla\alpha+\alpha\nabla_x V)=0$
and therefore $\alpha=Ce^{-V}$.
\end{proof}
\subsection{Existence of steady states}
The existence of steady states for the VPFP system in the gravitational case is not yet well-understood. We mention that a small mass result is proved in~\cite{bdol} for the VPFP system using a fixed point argument inspired by~\cite{dr,dr2}. This argument applies {\it mutatis mutandis} to the VNFP system: Consider the equation for the gravitational potential of steady states, eq.~\eqref{eliN}, which we rewrite in terms of $u=-\phi_0$ as
\begin{equation}\label{eliN2}
\Delta u=-\frac{e^{-V}Me^{-2u}}{\int_{\mathbb{R} ^6}\!e^{-\sqrt{e^{-2u}+|p|^2}-V}\,\mathrm{d}p\,\mathrm{d} x}\int_{\mathbb{R} ^3}\!\frac{e^{-\sqrt{e^{-2u}+|p|^2}}}{\sqrt{e^{-2u}+|p|^2}}\,\mathrm{d}p.
\end{equation}
Define by $K$ the solution operator of~\eqref{eliN2}, i.e., the convolution of the r.h.s. with $1/(4\pi |x|)$. By standard estimates one can prove that, for $M$ small enough, the operator $K$ is a contraction in the space $X=\{v\in L^\infty(\mathbb{R}^3):0\leq v\leq 1\}$ and so by the fixed point theorem we have the following result.
\begin{proposition}\label{smallM}
There exists $M_0>0$ such that, for all $M<M_0$, the equation~\eqref{eliN2}, with the boundary condition $\lim_{|x|\to\infty}u=0$, has a unique solution $u\in L^\infty$. This solution defines, through~\eqref{staticVNFP}, a steady state of the VNFP system.
\end{proposition}
However in the relativistic case we can do much better, and prove existence of steady states for {\it all} masses. Let us denote
\[
\Gamma_M=\{f:\mathbb{R}^6\to\mathbb{R}\,:\, f\in L^1(\mathbb{R}^6)\,,\, \|f\|_{L^1(\mathbb{R}^6)}=M\,,\,\mathcal{Q}_+[f]<\infty\},
\]
where $\mathcal{Q}_+$ is defined by~\eqref{q+},
and recall that the space $D^1(\mathbb{R}^3)$ is defined as
\[
D^1(\mathbb{R}^3)=\{\phi\in L^1_{\rm loc}(\mathbb{R}^3):\,\nabla\phi\in
L^2\textnormal{ and }\phi \textnormal{ vanishes at infinity}\},
\]
where the condition of $\phi$ vanishing at infinity means that the
set $\{x\in\mathbb{R}^3:|\phi(x)|>a\}$ has finite (Lebesgue) measure, for all
$a>0$.  Functions in the space $D^1(\mathbb{R}^3)$ satisfy the Sobolev inequality
\begin{equation}\label{sobolev}
\|\phi\|_{L^6}\leq
\eta\|\nabla\phi\|_{L^2},\quad\eta=\frac{2}{\sqrt{3}}\pi^{-2/3},
\end{equation}
see \cite[Thm.~8.3]{LL}.
\begin{theorem}
For all $M>0$ there exists at least one solution\footnote{By Proposition~\ref{smallM}, the solution is unique for $M$ small.} $\phi_0$ of~\eqref{eliN}. Moreover the corresponding steady state,  given by~\eqref{staticVNFP2}, is a minimizer of the entropy functional:
\[
K(M)=\inf\{\mathcal{K}(f,\phi,\psi)\,, f\in\Gamma_M\,,\,\phi\in D^1(\mathbb{R}^3)\,,\psi\in L^2(\mathbb{R}^3)\},
\]
where $\mathcal{K}$ is defined by~\eqref{eln}, i.e., $K(M)=\mathcal{K}(m_M,\phi_0,0)$.
\end{theorem}
\begin{proof}
First we notice that
\[
K(M)=\inf_{\Gamma_M\times D^1}\mathcal{E}(f,\phi),
\]
where $\mathcal{E}(f,\phi)=\mathcal{K}(f,\phi,0)$. We divide the proof in five steps.

{\it Step 1: $K(M)$ is bounded.} We have
\begin{equation}\label{boundbelow}
\mathcal{E}(f,\phi)\geq\int_{\mathbb{R}^6} \!f(|p|+V(x)+\log f)\,\mathrm{d}p\,\mathrm{d}x.
\end{equation}
Using Lemma~\ref{dol1} with $g=f$, $h=|p|+V$, $\Omega=\mathbb{R}^{6}$ we get
\[
\mathcal{E}(f,\phi)\geq M\log\left(\frac{M}{\int_{\mathbb{R}^6}\!e^{-|p|-V}\mathrm{d}p\,\mathrm{d}x}\right).
\]
{\it Step 2: Weak convergence of minimizing sequences.} Let $(f_n,\phi_n)$ be a minimizing sequence. Since $\phi_n$ is uniformly bounded in $D^1$, and by the Sobolev inequality~\eqref{sobolev},  there exists a subsequence, still denoted by $\phi_n$, and $\phi_0\in D^1$ such that
\begin{equation}\label{weakphin}
\phi_n{\rightharpoonup}\phi_0 \text{ in }  L^6 \text{ and } \nabla_x\phi_n\rightharpoonup\nabla_x\phi_0 \text{ in }L^2.
\end{equation}
Next we establish the weak convergence of $f_n$ in $L^1$ by using the argument in~\cite[pag.~129]{dol}. Let us show first that $f_n$ does not concentrate. If it did, we could find $\varepsilon>0$, a bounded sequence $x_n\in\mathbb{R}^3$ and a sequence $R_n\to\infty$ such that
\[
\int_{|x_n-x|\leq R_n}\!f_n(x,p)\,\mathrm{d}p\,\mathrm{d}x=\varepsilon,\quad \text{for all }n\in\mathbb{N}.
\]
From~\eqref{boundbelow} and Lemma~\ref{dol1} we have
\begin{align}
\mathcal{E}(f_n,\phi_n)\geq& \int_{|x-x_n|>R_n}\!f_n(\log f_n+|p|+V(x))\,\mathrm{d}p\,\mathrm{d}x\nonumber\\
&+\int_{|x_n-x|\leq R_n}\!f_n(\log f_n+|p|+V(x))\,\mathrm{d}p\,\mathrm{d}x\nonumber\\
\geq&(M-\varepsilon)\log\left(\frac{M-\varepsilon}{\int_{|x_n-x|> R_n}\!e^{-|p|-V}\mathrm{d}p\,\mathrm{d}x}\right)\nonumber\\
&+ \varepsilon\log\left(\frac{\varepsilon}{\int_{|x_n-x|\leq R_n}\!e^{-|p|-V}\mathrm{d}p\,\mathrm{d}x}\right).\label{split}
\end{align}
Since $e^{-|p|-V}\in L^1$, we have
\begin{align*}
&\lim_{n\to\infty}\int_{|x_n-x|> R_n}\!e^{-|p|-V}\mathrm{d}p\,\mathrm{d}x=0,\\
& \text{ and }\ \lim_{n\to\infty}\int_{|x_n-x|\leq R_n}\!e^{-|p|-V}\mathrm{d}p\,\mathrm{d}x=\|e^{-|p|-V}\|_{L^1(\mathbb{R}^6)}
\end{align*}
and so~\eqref{split} implies $\mathcal{E}(f_n,\phi_n)\to\infty$ as $n\to \infty$, which contradicts the fact that $(f_n,\phi_n)$ is a minimizing sequence. Now we prove that $f_n$ is tight. If not, we can find $\varepsilon>0$ such that, for all $R_0>0$, there exists $R>R_0$ such that
\[
\lim_{n\to\infty}\int_{|x|+|p|>R}\!f_n\,\mathrm{d}p\,\mathrm{d}x>\varepsilon.
\]
Whence, using again~\eqref{boundbelow} and Lemma~\ref{dol1},
\begin{align*}
\mathcal{E}(f_n,\phi_n)\geq \left(\int_{|x|+|p|>R}\!f_n\,\mathrm{d}p\,\mathrm{d}x\right)\Big[&\log\int_{|x|+|p|>R}\!f_n\,\mathrm{d}p\,\mathrm{d}x\\
& -\log\int_{|x|+|p|>R}\!e^{-|p|-V}\mathrm{d}p\,\mathrm{d}x\Big]
\end{align*}
and so
\[
\lim_{R\to\infty}\int_{|x|+|p|>R}\!e^{-|p|-V}\mathrm{d}p\,\mathrm{d}x\geq \varepsilon e^{-K(M)/\varepsilon}>0,
\]
which contradicts the fact that $e^{-|p|-V}\in L^1$. We conclude that there exists $f_0\in L^1$ and a subsequence $f_n$ such that
\begin{equation}\label{weakfn}
f_n\rightharpoonup f_0 \text{ in }L^1.
\end{equation}
{\it Step 3: Pointwise convergence of minimizing sequences.} As proved in~\cite[Cor.~8.7]{LL}, the weak convergence~\eqref{weakphin} implies that
\begin{equation}\label{strongphin}
\phi_n\to \phi_0,\quad\text{pointwise a.e.}
\end{equation}
again up to the extraction of a subsequence.  Moreover, by the argument used in~\cite[Lemma~5]{CSS}, we may assume that $\phi_n\leq 0$ almost everywhere. Next we show that $f_n$ converges pointwise a.e. (up to subsequences).  Given a minimizing sequence $(f_n,\phi_n)$, define
\[
m_n=\frac{M}{\int_{\mathbb{R}^6}\!e^{-\sqrt{e^{2\phi_n}+|p|^2}-V}\,\mathrm{d}p\,\mathrm{d}x}e^{-\sqrt{e^{2\phi_n}+|p|^2}-V}.
\]
By Lemma~\ref{dol1} we have
\[
\mathcal{E}(f_n,\phi_n)-\mathcal{E}(m_n,\phi_n)\geq\frac{1}{2}\int_{\mathbb{R}^6}\!(\sqrt{f_n}-\sqrt{m_n})^2\,\mathrm{d}p\,\mathrm{d}x.
\]
This implies from one hand that $(m_n,\phi_n)$ is again a minimizing sequence and, on the other hand, that $\lim_{n\to\infty}(f_n-m_n)=0$  pointwise a.e. after extracting a suitable subsequence. Moreover, since
\[
e^{-\sqrt{e^{2\phi_n}+|p|^2}-V}\to e^{-\sqrt{e^{2\phi_0}+|p|^2}-V},\quad\text{pointwise a.e.}
\]
and (by dominated convergence)
\[
\int_{\mathbb{R}^6}\! e^{-\sqrt{e^{2\phi_n}+|p|^2}-V}\,\mathrm{d}p\,\mathrm{d}x\to \int_{\mathbb{R}^6}\!e^{-\sqrt{e^{2\phi_0}+|p|^2}-V}\,\mathrm{d}p\,\mathrm{d}x,\quad\text{pointwise a.e.}
\]
then
\[
f_n\to \frac{M}{\int_{\mathbb{R}^6}\!e^{-\sqrt{e^{2\phi_0}+|p|^2}-V}\,\mathrm{d}p\,\mathrm{d}x}e^{-\sqrt{e^{2\phi_0}+|p|^2}-V},\quad\text{pointwise a.e.}
\]
In particular we notice that $f_0$ is strictly positive and bounded.

{\it Step 4: $(f_0,\phi_0)$ is a minimizer.} We prove that $\mathcal{E}$ is weakly lower semicontinuous. Clearly
\[
\liminf_{n\to\infty}\int\! |\nabla_x\phi_n|^2\,\mathrm{d}x\geq \int\!|\nabla_x\phi_0|^2\,\mathrm{d}x.
\]
Moreover, by Fatou's lemma,
\[
\liminf_{n\to\infty}\int_{\mathbb{R}^6}\!f_n(\sqrt{e^{2\phi_n}+|p|^2}+V+\log f_n)\,\mathrm{d}p\,\mathrm{d}x\geq\int_{\mathbb{R}^6}\!f_0(\sqrt{e^{2\phi_0}+|p|^2}+V+\log f_0),
\]
and the claim follows: $K(M)=\mathcal{E}(f_0,\phi_0)$.

{\it Step 5: $(f_0,\phi_0)$ is a steady state of the VNFP system.} Since we already proved in step 3 that
\[
f_0= \frac{M}{\int_{\mathbb{R}^6}\!e^{-\sqrt{e^{2\phi_0}+|p|^2}-V}\,\mathrm{d}p\,\mathrm{d}x}e^{-\sqrt{e^{2\phi_0}+|p|^2}-V},
\]
we only need to show that $\phi_0$ solves the non-linear elliptic equation~\eqref{eliN}. To this purpose we define $\phi_h=\phi_0+h\eta$, where $\eta=\eta(x)$ is any $C^\infty$ function with compact support and $h\in\mathbb{R}$. Using that $0<f_0<\infty$ and $\phi_0\leq 0$, it is straightforward to show that $\mathcal{E}(f_0,\phi_h)$ is differentiable in $h$. The derivative at $h=0$ must vanish and this entails that $\phi_0$ solves
\[
\Delta\phi_0=e^{2\phi_0}\int_{\mathbb{R}^3}\!\frac{f_0}{\sqrt{e^{2\phi_0}+|p|^2}}\,\mathrm{d}p\,\mathrm{d}x
\]
in the sense of distributions. This completes the proof of the theorem.
\end{proof}

\section*{Acknowledgments} This work was completed while the second author was a long term participant to the program ``Partial Differential Equations in Kinetic Theories" at the Isaac Newton Institute in Cambridge (UK). Support by  ``Ministerio Ciencia e Innovaci\'on", Spain (Project MTM2008-05271) is also acknowledged.

 \begin{appendix}
\section{The Cauchy problem for the Fokker-Planck equation on a Riemannian manifold}\label{global}
In this appendix we discuss the initial value problem for the equation~\eqref{FPRfin}. In fact we shall study the problem for a more general equation than~\eqref{FPRfin}, where we allow for a general (positive definite) diffusion matrix $D$, a general velocity field (with non-zero gradient) in the transport operator and a general friction potential. Precisely we shall consider the initial value problem for the following equation:
\begin{equation}\label{kineticFP3}
\partial_t h+v(p)\cdot\nabla_xh-\nabla_x V\cdot\nabla_ph=\Delta_p^{(g)} h+Wh,\quad t>0,\ x\in\mathbb{R}^d,\ p\in\mathbb{R}^d,
\end{equation}
where $\Delta_p^{(g)}$ denotes the Laplace-Beltrami operator of a Riemannian metric $g$ on $\mathbb{R}^d$, see~\eqref{LB}, and $W,v$ are the vector fields
\begin{equation}\label{W}
Wh=g^{-1}\nabla_p\log u\cdot\nabla_p h,\quad u=\sqrt{\det g}\,e^{-E},\quad v=\nabla_p E,
\end{equation}
for some non-negative function $E=E(p)$. Equation~\eqref{FPRfin} can be written in the form~\eqref{kineticFP3} by setting $g=D^{-1}$, $E(p)=\sqrt{1+|p|^2}$, $f=e^{-E-V}h$. We prove the following
\begin{theorem}\label{globalex}
Assume $g,E,V\in C^\infty$ satisfy $e^{-E},\, e^{-V}\in L^1(\mathbb{R}^3)$ and, for all $p\in\mathbb{R}^d$,
\begin{align}
&\det(\partial_{p^i}v_j)\neq 0, \nonumber\\
&\partial_{p^i}(g^{ij}\partial_{p^j}E)\leq \omega,\ \text{for some $\omega>0$},\nonumber\\
&g^{ij}\partial_{p^i} E\,\partial_{p^j} E\geq \theta|\nabla _p E|^2,\ \text{for some } \theta>0,
\label{metricnorm}\end{align}
where $g^{ij}$ is the inverse matrix of $g$, i.e., $g^{ij}g_{jk}=\delta_k^i$.
Furthermore we assume that
\begin{equation}\label{boundstemp}
\frac{g^{ij}(p)}{|p|^2}\to 0,\ \text{as }|p|\to\infty\ \ \forall\, i,j=1,\dots d.
\end{equation}
Then for all $h_\mathrm{in}\in C^1_c(\mathbb{R}^d\times\mathbb{R}^d)$ there exists a unique
\[
h\in C([0,\infty),L^2(\mathrm{d}\mu)),
\]
solution of~\eqref{kineticFP3} with initial datum $h_\mathrm{in}$, where $\mu$ is the probability measure defined by
\[
\mathrm{d}\mu=\Theta^{-1}e^{-E-V}\,\mathrm{d}{p}\,\mathrm{d}{x},\quad \Theta=\int_{\mathbb{R}^{2d}}\!e^{-E(p)-V(x)}\,\mathrm{d}p\,\mathrm{d} x.
\]
\end{theorem}

For the proof we need the following lemma.

\begin{lemma}\label{aux}
Let $A=-L+T$, where $L=\Delta_p^{(g)} +W$ and $T=v(p)\cdot\nabla_x-\nabla_x V\cdot\nabla_p$, and $f,f_1,f_2\in C^\infty$. Then
\begin{itemize}
	\item[(a)] $\int_{\mathbb{R}^d} \!hLh\:e^{-E}\,\mathrm{d} p=-\int_{\mathbb{R}^d} \!g^{ij}\partial_{p^i}h\,\partial_{p^j} h\:e^{-E}\, \mathrm{d} p$;
	\item[(b)] $\int_{\mathbb{R}^{2d}}\! hTh\:\mathrm{d}{\mu}=0$;
	\item[(c)]  $A(f_1f_2)=f_1Af_2+f_2Af_1-2g^{ij}\partial_{p^i}f_1\,\partial _{p^j}f_2$.
\end{itemize}
\end{lemma}

\begin{proof}

The proof of (a) can be found in \cite{C}. For the second statement we use that
\begin{align*}
&\int_{\mathbb{R}^{2d}}\! h(v(p)\cdot\nabla_xh-\nabla_x V\cdot\nabla_ph)\:e^{-E-V}\,\mathrm{d} p\,\mathrm{d} x\\
&\qquad=\int_{\mathbb{R}^{2d}}\!( -\nabla_xh\cdot \nabla_p E\:h+\nabla_ph\cdot\nabla_x V)h\:e^{-E-V}\,\mathrm{d} p\,\mathrm{d} x,
\end{align*}
using that $v(p)=\nabla_p E$ and integrating by parts.
The proof of (c) follows by Leibnitz's rule.
\end{proof}

\begin{proof}[Proof of Theorem~\ref{globalex}]
We generalize the proof of~\cite[Prop.~5.5]{hn}, where the following argument is applied to the non-relativistic Fokker-Planck equation~\eqref{FPequation} (with external potential), and the proof in~\cite[App.~A]{C}, which studies the Cauchy problem for~\eqref{kineticFP3} when $x\in\mathbb{T}^d$ (the $d-$dimensional torus) without external potential.

Denote $\mathcal{H}=L^2(\mathrm{d}\mu)$. Let us consider the operator
\[
A=v(p)\cdot\nabla_xh-\nabla_x V\cdot\nabla_ph-\Delta_p^{(g)} h-Wh=T-L
\]
defined on $D(A)=C^\infty_c(\mathbb{R}^{2d})$. Equation~\eqref{kineticFP3} takes the form $\partial_th+A h=0$. Our goal is to show that the closure of the operator $A$ generates a contraction semigroup on $\mathcal{H}$. To this purpose it suffices to prove that $A$ is accretive and that the range of $A+\lambda I$ is  dense in $\mathcal{H}$ for some $\lambda>0$, see~\cite[Sec.~5.2]{hn}.

That $A$ is accretive follows by (a) and (b) of the previous lemma:
\[
\left\langle h\left|\right. Ah\right\rangle_\mathcal{H}=-\left\langle h\left|\right. Lh\right\rangle_\mathcal{H}+\left\langle h\left|\right. Th\right\rangle_\mathcal{H}= \int\! g^{ij}\partial_{p^i} h\, \partial_{p^j} h\,\mathrm{d}\mu \geq 0.
\]

Next we show that $A$ is hypoelliptic. Let $a=\sqrt{g^{-1}}$, the positive definite matrix such that $a^2=g^{-1}$. A direct computation shows that
\[
-A=\sum_{i=1}^d Y^2_{i}+Y_0,
\]
where $Y_0,Y_i$ denote the vector fields
\begin{align*}
Y_0h&=(\textrm{div}_p a)\cdot a\nabla_p h-g^{ij}\partial_{p^i} E\,\partial_{p^j} h-Th,\\
Y_{i}h&=a_i^k\partial_{p^k}h.
\end{align*}

In order to prove that $A$ is hypoelliptic, it is enough to show that $-A$ satisfies a rank 2 Hormander's condition, i.e., the vector fields $Y_i$ and $Z_i:= [Y_0,Y_i]$ form a basis of $\mathbb{R}^{2d}$, see~\cite{H}. Observe that
\[
Z_i= B^k_i\partial_{p^k}+C^j_i\partial_{x^j},
\]
where $C_i^j=a_i^k\partial_{p^k}v^j$ and $B$ is a $d\times d$ matrix whose exact form is irrelevant for what follows. Thus we can represent the linear transformation $\{\partial_{x^i}, \partial_{p^j}\}\to \{Y_k, Z_l\}$ by
\[
F=\left( \begin{array}{cc}
0 & a\\
C & B
\end{array} \right),
\]
whose determinat is $|\!\det F|=\det a|\!\det C|= \det g|\!\det(\partial_{p^k}v^j)|$, which is positive because $\det(\partial_{p^i}v_j)$ is non-zero by assumption. Therefore, $\{Y_i, Z_j\}$ is a basis of $\mathbb{R}^{2d}$.\\

Finally, we prove that the range of $\lambda +A$ is dense in $\mathcal{H}$ for some $\lambda>0$. If $h\in \mathcal{H}$, we must show that if
\begin{equation}\label{fullrange}
\left\langle h|(\lambda +A)f\right\rangle_{\mathcal{H}}=0,\quad \textrm{for all } f \in D(A),
\end{equation}
then $h=0$. Equation~\eqref{fullrange} is equivalent to $h$ being a distributional solution of
\[
(\lambda -L-T)h=0.
\]
Since the operator in the left hand side of the latter equation is hypoelliptic, then we can assume $h\in C^\infty$. Now setting $f_1=\phi$, $f_2=\phi h$ in (c) of Lemma~\ref{aux}, multiplying by $h$, integrating and using that $\left\langle h|(\lambda +A)(\phi^2 h)\right\rangle_\mathcal{H}=0$, by~\eqref{fullrange}, we obtain
\begin{equation}\label{identity}
\lambda \int\! \phi ^2h^2\,\mathrm{d}\mu+\int \!g^{ij}\partial_{p^i}(\phi h)\,\partial_{p^j}(\phi h)\,\mathrm{d}\mu=\int \!h^2g^{ij}\partial_{p^i}\phi\,\partial_{p^j}\phi\,\mathrm{d}\mu-\int\! h^2\phi T\phi\,\mathrm{d}\mu.
\end{equation}
Setting $f=he^{-E/2-V/2}$ we get
\begin{align*}
\int\! g^{ij}\partial_{p^i}(\phi h)\,\partial_{p^j}(\phi h)\,\mathrm{d} \mu =&\int\! g^{ij}\partial_{p^i}(\phi f)\,\partial_{p^j}(\phi f)\Theta^{-1}\,\mathrm{d} p\,\mathrm{d} x\\
& + \frac{1}{4}\int\! \phi^2 h^2g^{ij}\partial_{p^i} E\,\partial_{p^j} E\,\mathrm{d} \mu\\
&+ \frac{1}{2}\int\! g^{ij}\partial_{p^i}(\phi^2 f^2)\,\partial_{p^j} E\,\Theta^{-1}\,\mathrm{d} p\,\mathrm{d} x
\end{align*}
and integrating by parts in the last term we get
\[
\frac{1}{2}\int\! g^{ij}\partial_{p^i}(\phi^2 f^2)\,\partial_{p^j} E\,\Theta^{-1}\,\mathrm{d} p\,\mathrm{d} x=-\frac{1}{2}\int\! \phi^2 h^2\:\partial_{p^{i}}(g^{ij}\partial_{p^{j}}E)\,\mathrm{d} \mu\geq -\omega\int\! \phi^2 h^2\,\mathrm{d} \mu.
\]
The identity~\eqref{identity} leads therefore to the inequality
\[
(\lambda-\omega)\int\! \phi^2h^2\,\mathrm{d}\mu+\frac{1}{4}\int\! \phi^2 h^2g^{ij}\partial_{p^i} E\,\partial_{p^j} E\,d\mu \leq\int\! h^2g^{ij}\partial_{p^i}\phi\,\partial_{p^j}\phi\,\mathrm{d}\mu-\int\! h^2\phi T\phi\,\mathrm{d}\mu.
\]
Let $k=(k_1,k_2)\in \mathbb{N}^2$ and $\phi=\phi_k(x,p)=\psi(x/k_1)\psi(p/k_2)$, where $\psi\in C_c^\infty$, $0\leq \psi\leq 1$, $\psi=1$ on $B(0,1/2)$ and $\textrm{supp }\psi\subset B(0,1)$. We obtain, denoting by $C$ any positive constant,
\begin{align}
(\lambda -\omega) \int\! \phi_k^2h^2\,\mathrm{d}\mu + \frac{1}{4}\int\! \phi_k^2 h^2g^{ij}\partial_{p^i} E\,\partial_{p^j} E\,\mathrm{d}\mu\leq &\frac{C}{k_2^2}\int\! h^2\sup_{i,j}|g^{ij}|\chi_{|p|<k_2}\,\mathrm{d}\mu\nonumber\\
&+
|\left\langle\phi_k h\: \nabla_p E\cdot \nabla_x{\phi_k},h\right\rangle_\mathcal{H}|\nonumber\\
&+|\left\langle\phi_k h\nabla_x V \cdot\nabla_p{\phi_k},h\right\rangle_\mathcal{H}|.\label{inequalitytemp}
\end{align}
Using Young's inequality, we can estimate the last two terms of~\eqref{inequalitytemp} as
\begin{align*}
|\left\langle \phi_k h\:\nabla _p E\cdot \nabla_x{\phi_k},h\right\rangle| &  \leq \frac{C}{k_1}\left(\frac{1}{4\epsilon_1}\int\! \phi_k^2h^2|\nabla_p E|^2\mathrm{d}\mu+\epsilon_1\int\! h^2\mathrm{d}\mu\right),\\
|\left\langle \phi_k h\nabla_x V \cdot\nabla_p{\phi_k},h\right\rangle|
&\leq \frac{C \zeta(k_1)}{k_2}\left(\frac{1}{4\epsilon_2}\int\! \phi_k^2h^2\mathrm{d}\mu+\epsilon_2\int\! h^2\mathrm{d}\mu\right),
\end{align*}
for all $\epsilon_1,\epsilon_2>0$, where  $\zeta(k_1)=\sup_{|x|\leq k_1}\{|\nabla_x V|\}$. Taking $\epsilon _1= C/(\theta k_1)$ in the first line, $\epsilon _2= {C \zeta(k_1)}/(4k_2)$ in the second line and using (\ref{metricnorm}), we get
 \begin{align*}
(\lambda -\omega-1) \int\! \phi_k^2h^2\mathrm{d}\mu \leq \frac{C}{k_2^2}\int\! h^2\sup_{i,j}|g^{ij}|\,\chi_{|p|<k_2}\mathrm{d}\mu+C\left(\frac{1}{k_1^2}+\frac{ C^2_{k_1}}{k_2^2}\right)\int\! h^2\mathrm{d}\mu.
\end{align*}
We see that $h=0$, taking first the limit  $k_2\to \infty$ and then $k_1\to \infty$. This concludes the proof of the theorem.
\end{proof}

\section{Finite propagation speed of relativistic kinetic equations}\label{finitespeedapp}
This appendix is devoted to prove a general result that can be used to establish the finite propagation speed property for all relevant relativistic kinetic equations. It is obtained by adapting the proof of a celebrated uniqueness theorem for non-linear wave equations due to Fritz John~\cite{FJ}, see also~\cite{So}.

\begin{lemma}\label{finitespeedlemma}
Let $\rho,j\in C^1$ verify
\begin{equation}\label{conteq2}
\partial_t\rho+\nabla\cdot j=0,\quad t\geq 0,\quad x\in\mathbb{R}^d,
\end{equation}
and $|j|\leq \rho$. If $\rho(0,x)=0$, for $|x-x_0|\leq t_0$, then $\rho(t,x)=0$, for $(t,x)\in\Lambda(t_0,x_0)$, where
\[
\Lambda(t_0,x_0)=\{(t,x)\in[0,t_0]\times \mathbb{R}^d:|x-x_0|\leq t_0-t\}.
\]
\end{lemma}
\begin{proof}
Consider the function
\[
\Phi(s,x)=t_0-[(t_0-s)^2+t_0^{-2}(2t_0s-s^2)|x-x_0|^2]^{1/2}.
\]
Note that
\begin{equation}\label{prophi}
\Phi(0,x)=0,\quad\lim_{s\to t_0}\Phi(s,x)=t_0-|x-x_0|,\quad\text{and}\quad \Phi_{|_{|x-x_0|=t_0}}=0.
\end{equation}
Moreover, denoting $R_s(t_0,x_0)=\{(t,x):t\leq\Phi(s,x),\ |x-x_0|\leq t_0\}$, we have
\[
\Lambda(t_0,x_0)=\cup_{0\leq s<t_0}R_s(t_0,x_0).
\]
Next we define
\[
\rho_{\cap}(s,x)=\rho(\Phi(s,x),x),\quad j_{\cap}(s,x)=j(\Phi(s,x),x).
\]
Since $\rho,j$ satisfy~\eqref{conteq2}, then $\rho_{\cap}$, $j_{\cap}$ verify
\[
\partial_s\rho_{\cap}=-\nabla\cdot j_{\cap}\partial_s\Phi+\partial_sj_{\cap}\cdot\nabla_x\phi.
\]
Therefore, using~\eqref{prophi},
\begin{align*}
\int_{|x-x_0|<t_0}\!\rho_{\cap}(s,x)\,\mathrm{d}x&=\int_{|x-x_0|<t_0}\int_0^s\!\partial_\tau\rho_{\cap}(\tau,x)\,\mathrm{d}\tau\,\mathrm{d}x\\
&=\int_{|x-x_0|<t_0}\int_0^s\!(-\nabla\cdot j_{\cap}\partial_\tau\Phi+\partial_\tau j_{\cap}\cdot\nabla_x\phi)\,\mathrm{d}\tau\,\mathrm{d}x\\
&=-\int_{|x-x_0|<t_0}\!\nabla\cdot j_{\cap}\,\Phi(s,x)\,\mathrm{d}x\\
&=\int_{|x-x_0|<t_0}\!j_{\cap}\cdot\nabla_x\phi(s,x)\,\mathrm{d}x,
\end{align*}
whence
\[
\int_{|x-x_0|<t_0}\!(\rho_{\cap}-j_{\cap}\cdot\nabla_x\phi)\,\mathrm{d}x=0.
\]
Moreover it is easy to verify that $|\nabla_x\phi(s,x)|\leq\theta(s)<1$, for all $0\leq s<t_0$, thus, since in addition $|j|\leq \rho$, we get
\[
\int_{|x-x_0|<t_0}\!\rho_{\cap}\,\mathrm{d}x=0\Rightarrow\rho=0\ \text{on }\ \Lambda(t_0,x_0).
\]
\end{proof}
The preceding lemma can be applied to any relativistic kinetic equation which is compatible with the continuity equation~\eqref{conteq2}. Precisely, to any kinetic equation of the form
\[
\partial_tf+\hat{p}\cdot\nabla_x f=Q[f],
\]
where $Q$ is a (possibly non-linear) operator such that
\[
\int_{\mathbb{R}^d}\!Q[f](t,x,p)\mathrm{d}p=0.
\]
The previous identity implies that
\[
\rho=\int_{\mathbb{R}^d} \!f\,\mathrm{d}p,\quad j=\int_{\mathbb{R}^d}\!f\,\hat{p}\,\mathrm{d}p
\]
satisfy the continuity equation~\eqref{conteq2}. Moreover, since $|\hat{p}|\leq 1$, then $|j|\leq\rho$ and Lemma~\ref{finitespeedlemma} applies.
\end{appendix}

% You may incorporate your references as follows in your main tex file.
% Using BibTex is not recommended but can be handled.

\end{document}